\documentclass[preprint2]{aastex}  

\usepackage{amsmath,times,ulem}
\usepackage[displaymath]{lineno}
\usepackage[OT1,T1]{fontenc} 

\shorttitle{Solar Magnetic Field Reversals}

\shortauthors{DeRosa et al.}

\newcommand{\epsl}{Earth Planet.\ Sci.\ Lett.}
\newcommand{\lrsp}{Liv.\ Rev.\ Sol.\ Phys.}
\newcommand{\pplas}{Phys.\ Plasmas}

\newcommand{\pepi}{PEPI}
\newcommand{\anach}{Astron.\ Nachr.}
\newcommand{\gafd}{Geophys. Astrophys. Fluid Dyn.}
\newcommand{\geojint}{Geophys. J. Int.}
\newcommand{\pnas}{PNAS}
\newcommand{\rsta}{Phil. Trans. R. Soc. Lond. A}

\newcommand{\bs}{\boldsymbol} 
\newcommand{\bcr}{\bs\times} 
\newcommand{\bdel}{\bs\nabla} 
\newcommand{\bdot}{\bs\cdot} 
\newcommand{\delcr}{\bdel\bcr} 
\newcommand{\deldot}{\bdel\bdot} 

\begin{document}

\renewcommand{\textfraction}{0.1}
\renewcommand{\floatpagefraction}{0.2}
\renewcommand{\dblfloatpagefraction}{0.7}
\renewcommand{\dbltopfraction}{0.7}


\title{Solar Magnetic Field Reversals and the Role of Dynamo Families}

\author{M.L.~DeRosa\altaffilmark{1}, A.S.~Brun\altaffilmark{2}, and J.T.~Hoeksema\altaffilmark{3}}

\altaffiltext{1}{Lockheed Martin Solar and Astrophysics
Laboratory, 3251 Hanover St. B/252, Palo Alto, CA 94304, USA}

\altaffiltext{2}{Laboratoire AIM Paris-Saclay, CEA/Irfu Universit\'e
  Paris-Diderot CNRS/INSU, 91191 Gif-sur-Yvette, France}

\altaffiltext{3}{W.~W.~Hansen Experimental Physics Laboratory, Stanford
  University, Stanford, CA 94305, USA}

\begin{abstract}

The variable magnetic field of the solar photosphere exhibits periodic
reversals as a result of dynamo activity occurring within the solar interior.
We decompose the surface field as observed by both the Wilcox Solar
Observatory and the Michelson Doppler Imager into its harmonic constituents,
and present the time evolution of the mode coefficients for the past three
sunspot cycles.  The interplay between the various modes is then interpreted
from the perspective of general dynamo theory, where the coupling between the
primary and secondary families of modes is found to correlate with large-scale
polarity reversals for many examples of cyclic dynamos.  Mean-field dynamos
based on the solar parameter regime are then used to explore how such
couplings may result in the various long-term trends in the surface magnetic
field observed to occur in the solar case.

\end{abstract}


\section{Introduction} \label{sec:intro}

The Sun is a dynamic star that possesses quasi-regular cycles of magnetic
activity having a mean period of about 22~yr.  This period varies from cycle
to cycle, and over the past several centuries has ranged from 18 to 25 years
\citep{wei1990,bee1998,uso2007}, as for example illustrated by the unusual but
not unprecedented length of the most recently completed sunspot cycle~23.
During each sunspot cycle (comprising half of a magnetic cycle), the Sun
emerges sunspot groups and active regions onto the photosphere, with such
features possessing characteristic latitudes, polarity, and tilt angles.  As
with the period, the numbers and emergence frequencies of active regions is
observed to vary from cycle to cycle.

At activity minima when few active regions are present, the surface magnetic
field is characterized by the presence of two polar caps, i.e., largely
unipolar patches of magnetic flux dispersed across both polar regions with the
northern and southern caps possessing opposite polarities.  Reversals of this
large-scale dipole represented by the polar-cap flux occur during each sunspot
cycle, allowing the subsequent sunspot cycle to begin in the opposite
configuration.  After two sunspot cycles, and thus after undergoing two
polarity reversals, the photospheric field will have returned to its starting
configuration so as to complete a full activity cycle.

In response to the photospheric flux associated with various features, such as
active regions and their decay products, the coronal magnetic field possesses
structures having a broad spectrum of sizes.  These structures are both
evident in observations of coronal loops, as found in narrow-band extreme
ultraviolet or soft X-ray imagery, and reproduced in models of the coronal
magnetic field (e.g., \citealt{sch2003}).  In both venues, the coronal
magnetic field geometry is seen to contain a rich and complex geometry.
Dynamical events originating from the corona, such as eruptive flares and
coronal mass ejections, are likely powered by energy released by a
reconfiguration of the coronal magnetic field, which in turn is responding to
changes and evolution of photospheric fields.

Precise measurements of the time-history of photospheric magnetic field and
the ability to determine the projection of this field into its constituent
multipole components are helpful in investigating the physical processes
thought to be responsible for dynamo activity \citep{bul1954,mof1978}.  In
cool stars similar to the Sun, the dynamo is presumed to be a consequence of
the nonlinear interactions between convection, rotation, and large scale
flows, leading to the generation and maintenance against Ohmic diffusion of
magnetic field of various temporal and spatial scales
\citep{wei1987,cat1999,oss2003,bru2004,voe2007,cha2010,rei2012}.  In
particular, the dependence of dynamo activity upon rotation appears to be well
established \citep{noy1984,saa1999,piz2003,boh2007,rei2009}.  However, many
details of the understanding of why many cool-star dynamos excite waves of
dynamo activity having a regular period, specifically 22~yr in the case of the
Sun, remain unclear.

To investigate this question, it is useful to explore the behavior and
evolution of the lowest-degree (i.e., largest-scale) multipoles, their
amplitudes and phases, and their correlations with the solar photospheric
magnetic field.  Many earlier studies (e.g.,
\citealt{lev1977b,hoe1984,gok1992a,gok1992b}) have illustrated how power in
these modes ebb and flow as a function of the activity level.  In particular,
J.~Stenflo and collaborators have performed thorough spectral analyses on the
temporal evolution of the various spherical harmonic modes.  \citet{ste1986},
\citet{ste1987}, and \citet{ste1988}, and more recently \citet{ste1994} and
\citet{kna2005}, base their analysis on Mt.~Wilson and Kitt Peak magnetic data
spanning the past few sunspot cycles.  As one would expect, they find that
most of the power is contained in temporal modes having a period of about
22~yr, and especially in spherical harmonics that are equatorially
antisymmetric, such as the axial dipole and octupole.  However, they find
signatures of the activity cycle are present in all axisymmetric harmonics, as
significant power is present at temporal frequencies at or near integer
multiples of the fundamental frequency of 1.44~nHz [equivalent to
  (22~yr)$^{-1}$].

In the current study, we focus on the coupling between spherical harmonic
modes, and what such coupling may indicate about the operation of the interior
dynamo.  In particular, reversals of the axial dipole mode may be viewed as a
result of continuous interactions between the poloidal and toroidal components
of the interior magnetic field, i.e., the so-called dynamo loop.  Currently,
one type of solar dynamo model that successfully reproduces many observed
behaviors is the flux-transport Babcock-Leighton type (e.g.,
\citealt{cho1995,dik2004,jou2007,yea2008}).  A key ingredient in producing
realistic activity cycles using this type of model is found to be the
amplitude and profile of the meridional flow
\citep{jou2007,kar2010,nan2011,dik2011}, which result in field reversals
progressing via the poleward advection across the surface of trailing-polarity
flux from emergent bipolar regions.  During the rising phase of each sunspot
cycle, polar cap flux left over from the previous cycle is canceled, after
which new polar caps having the opposite magnetic polarity form
\citep{wany1989,ben2004,das2010}.

Helioseismic analyses of solar oscillations have provided measurements and
inferences of key dynamo components, such as the internal rotation profile and
the near-surface meridional circulation \citep{tho2003,bas2010}.
Complementing precise observations of the solar magnetic cycle properties,
these helioseismic inversions represent additional strong constraints on
theoretical solar dynamo models.  Successful solar dynamo models strive to
reproduce as many empirical features of solar magnetic activity as possible,
including not only cycle periods, but also parity, phase relation between
poloidal and toroidal components, and the phase relation between the dipole
and higher-degree harmonic modes.

Interestingly, a recent analysis of geomagnetic records has indicated that the
interplay between low-degree harmonic modes during polarity reversals is one
way to characterize both reversals of the geomagnetic dynamo (which have a
mean period of about 300,000~yr) as well as excursions, where the dipole axis
temporarily moves equatorward and thus away from its usual position of being
approximately aligned with the rotation axis, followed by a return to its
original position without having crossed the equator (see \citealt{hul2010}
for a recent review on Earth's magnetic field).  In particular, these studies
have shown that, during periods of geomagnetic reversals, the quadrupolar
component of the geomagnetic field is stronger than the dipolar component,
while during an excursion (which can be thought of as a failed reversal), the
dipole remains dominant \citep{ami2010,leo2007,leo2009}.  One may thus ask: Is
a similar behavior observed for the solar magnetic field?

In an attempt to address this question, we have performed a systematic study
of the temporal evolution of the solar photospheric field by determining the
spherical harmonic coefficients for the photospheric magnetic field throughout
the past three sunspot cycles, focusing on low-degree modes and the relative
amplitude of dipolar and quadrupolar components.  Following the classification
of \citet{mcf1991}, we have made the distinction between primary and secondary
families of harmonic modes, a classification scheme that takes into account
the symmetry and parity of the spherical harmonic functions (see
\citealt{gub1993} for a detailed discussion on symmetry and dynamo).

While we recognize that the solar dynamo operates in a more turbulent
parameter regime than the geodynamo, and is more regular in its reversals, the
presence of grand minima (such as the Maunder Minimum) in the historical
record indicates that the solar dynamo can switch to a more intermittent state
on longer-term, secular time scales.  In fact in the late stages of the
Maunder Minimum, the solar dynamo was apparently asymmetric, with the southern
hemisphere possessing more activity than the north \citep{rib1993} for several
decades, a magnetic configuration that may have been achieved by having
dipolar and quadrupolar modes of similar amplitude \citep{tob1997,gal2009}.
Additionally, recent spectopolarimetric observations of solar-like stars now
provide sufficient resolution to characterize the magnetic field geometry in
terms of its multipolar decomposition \citep{pet2008b}.  Furthermore, the
analysis of reduced dynamical systems developed over the last 20~yr describing
the geodynamo and solar dynamo have emphasized the importance of the nonlinear
coupling between dipolar and quadrupolar components
\citep{kno1996,wei2000,pet2009}.

This article is organized in the following manner.  In
\S\ref{sec:observations}, we describe the data sets and the data analysis
methods used to perform the spherical harmonic analysis, followed in
\S\ref{sec:expansion} with an explanation of the temporal evolution of the
various harmonic modes, the magnetic energy spectra, and the decomposition in
terms of primary and secondary families.  We interpret in
\S\ref{sec:discussions} our results from a dynamical systems perspective and
illustrate some of these concepts using mean-field dynamo models.  Concluding
remarks are presented in \S\ref{sec:conclusions}.

\section{Observations and Data Processing} \label{sec:observations}

We analyze time series of synoptic
photospheric magnetic field maps of the radial magnetic field $B_r$ derived
from line-of-sight magnetogram observations taken by both the Wilcox Solar
Observatory (WSO; \citealt{sch1977}) at Stanford University and by the
Michelson Doppler Imager (MDI; \citealt{sch1995}) on board the space-borne
Solar and Heliospheric Observatory (SOHO).  The WSO data\footnote{Available at
  \url{http://wso.stanford.edu/synopticl.html}.}  used in this study span the
past 36 years, commencing with Carrington Rotation (CR) 1642 (which began on
1976~May~27) and ending with CR~2123 (which ended on 2012~May~25).  For
MDI\footnote{Available at
  \url{http://soi.stanford.edu/magnetic/index6.html}.}, we used data from much
of its mission lifetime, starting with CR~1910 (which began on 1996~Jul.~1)
through CR~2104 (which ended on 2010~Dec.~24).  In both data series, one map
per Carrington rotation was used, though maps with significant amounts of
missing data were excluded.  The measured line-of-sight component of the field
is assumed to be the consequence of a purely radial magnetic field when
calculating the harmonic coefficients.  Additionally, for WSO, the synoptic
map data are known to be a factor of about 1.8 too low due to the saturation
of the instrument \citep{sva1978}.  Lastly, the MDI data have had corrections
applied for the polar fields using the interpolation scheme presented in
\citet{sun2011}.

For each map, we perform the harmonic analysis using the Legendre-transform
software provided by the ``PFSS'' package available through SolarSoft.  Using
this software first entails remapping the latitudinal dimension of the input
data from the sine-latitude format provided by the observatories onto a
Gauss-Legendre grid (c.f., \S~25.4.29 of \citealt{abr1972}).  This regridding
enables Gaussian quadrature to be used when evaluating the sums needed to
project the magnetic maps onto the spherical harmonic functions.  The end
result is a time-varying set of complex coefficients $B_{\ell}^m(t)$ for a
series of modes spanning harmonic degrees $\ell$ = 0, 1, ..., $\ell_{\max}$,
where the truncation limit $\ell_{\max}$ is equal to 60 for the WSO maps and
192 for MDI maps.  The $B_\ell^m$ coefficients are proportional to the
amplitude of each spherical harmonic mode $Y_{\ell}^m$ for degree $\ell$ and
order $m$ possessed by the time series of synoptic maps, so that
\begin{equation}
  B_r(\theta,\phi,t) = \sum_{\ell=0}^{\ell_{\max}} \sum_{m=0}^{\ell}
  B_\ell^m(t)\, Y_{\ell}^m (\theta,\phi),
  \label{eq:expansion}
\end{equation}
where $\theta$ is the colatitude, $\phi$ is the latitude, and $t$ is time.  We
note that because the coefficients $B_\ell^m$ are complex numbers, this
naturally accounts for the rotational symmetry between spherical harmonic
modes with orders $m$ and $-m$ (for a given value of $\ell$), with the
amplitudes of modes for which $m>0$ appearing in the real part of $B_\ell^m$,
and the amplitudes of the modes where $m<0$ being contained in the imaginary
part of $B_\ell^m$.  Consequently, the sum over $m$ in
equation~(\ref{eq:expansion}) starts at $m=0$ instead of at $m=-\ell$.  The
coefficients $B_\ell^0$ corresponding to the axisymmetric modes (for which
$m=0$) are real for all $\ell$.

We use the convention that, for a particular spherical harmonic degree $\ell$
and order $m$,
\begin{equation}
  Y_\ell^m(\theta,\phi) = C_\ell^m P_\ell^m (\cos\theta) \, e^{im\phi},
  \label{eq:shdef}
\end{equation}
where the functions $P_\ell^m(\cos\theta)$ are the associated Legendre
polynomials, and where the coefficients $C_\ell^m$ are defined
\begin{equation}
  C_\ell^m = (-1)^m \left[ \frac{2\ell+1}{4\pi}
  \frac{(\ell-m)!}{(\ell+m)!} \right]^\frac{1}{2}.  \label{eq:cdef}
\end{equation}
With this normalization, the spherical harmonic functions satisfy the
orthogonality relationship
\begin{equation}
  \int_0^{2\pi} d\phi \int_0^\pi \sin\theta \, d\theta \, Y_{\ell}^{m*}
  Y_{\ell'}^{m'} = \delta_{\ell\ell'} \, \delta_{mm'}.  \label{shortho}
\end{equation}
When comparing our coefficients with those from other studies, it is important
to take the normalization into account.  For example, the complex $B_{\ell}^m$
coefficients used here are different than (albeit related to) the real-valued
$g_{\ell}^m$ and $h_{\ell}^m$ coefficients provided by the WSO
team\footnote{Tables of $g_{\ell}^m$ and $h_{\ell}^m$ are available from the
  WSO webpage at
  \url{http://wso.stanford.edu/Harmonic.rad/ghlist.html}.}.
This difference is due to their use of spherical harmonics having the Schmidt
normalization, a convention that is commonly used by the geomagnetic community
as well as by earlier studies in the solar community such as \citet{alt1969}.
For the WSO data used here, we have verified that the values of $B_\ell^m$
used in this study are commensurate with the $g_\ell^m$ and $h_\ell^m$
coefficients provided by the WSO team.

Because we possess perfect knowledge of $B_r$ neither over the entire Sun nor
at one instant in time, the monopole coefficient function $B_0^0(t)$ does not
strictly vanish and instead fluctuates around zero.  In practice, we find that
the magnitude of $B_0^0(t)$ is small, and thus feel justified in not
considering it further.  This assumption effectively means that from each
magnetic map we are subtracting off any excess net flux, $\oint
B_r(\theta,\phi)\,\sin\theta\, d\theta\, d\phi$, a practice which leads to the
introduction of small errors in the resulting analysis.  However, these errors
are deemed to be less important than the inaccuracies resulting from the
less-than-perfect knowledge of the radial magnetic flux on the Sun, including
effects due to evolution and temporal sampling throughout each Carrington
rotation and due to the lack of good radial field measurements of the flux in
the polar regions of the Sun.

\section{Multipolar Expansions and Their Evolution as a Function of Cycle} \label{sec:expansion}

\subsection{Dipole Field (Modes with $\bs{\ell}=1$)}

The solar dipolar magnetic field can be analyzed in terms of its axial and
equatorial harmonic components.  As has long been known \citep{hoe1984}, the
axial dipole component, having a magnitude of $|B_1^0|$, is observed to be
largest during solar minimum when there is a significant amount of magnetic
flux located at high heliographic latitudes on the Sun.  These two so-called
\textsl{polar caps} possess opposite polarity, and match the polarity of the
trailing flux within active regions located in the corresponding hemisphere
that emerged during the previous sunspot cycle.  Long-term observations of
surface-flux evolution indicate that a net residual amount of such
trailing-polarity flux breaks off from decaying active regions and is released
into the surrounding, mixed-polarity quiet-sun network.  This flux is observed
to continually evolve as flux elements merge, fragment, and move around in
response to convective motions \citep{sch1997a}, but the long-term effect is
that the net residual flux is slowly advected poleward by surface meridional
flows.  Such poleward advection results in a net influx of trailing-polarity
flux into the higher latitudes.  At the same time, an equivalent amount of
leading-polarity flux from each hemisphere cancels across the equator, as is
necessary to balance the trailing-polarity flux advected poleward.  Over the
course of a sunspot cycle, this process is repeated throughout subsequent
sunspot cycles, during which flux from the trailing polarities of active
regions eventually cancels out the polar-cap flux left over from previous
cycles.  Once the leftover flux has fully disappeared, the buildup of a new
polar cap having the opposite polarity occurs by the subsequent activity
minimum.

In contrast to the axial dipole component, the equatorial dipole components,
having magnitudes $|B_1^{-1}|$ and $|B_1^1|$, are largest during maximum
activity intervals and weakest during activity minima.  Individual active
regions on the photosphere each contribute a small dipole moment that, aside
from the small axial component arising from the Joy's Law tilt, is oriented in
the equatorial plane.  Together the equatorial dipole moments from the
collection of active regions add vectorially to form the overall dipole
moment.  When many active regions are on the disk, it thus follows that the
equatorial dipole mode is likely to have a higher amplitude.  During periods
of quiet activity with few active regions on disk, the equatorial dipole
amplitude is minimal.

\begin{figure}
  \epsscale{1.0}
  \plotone{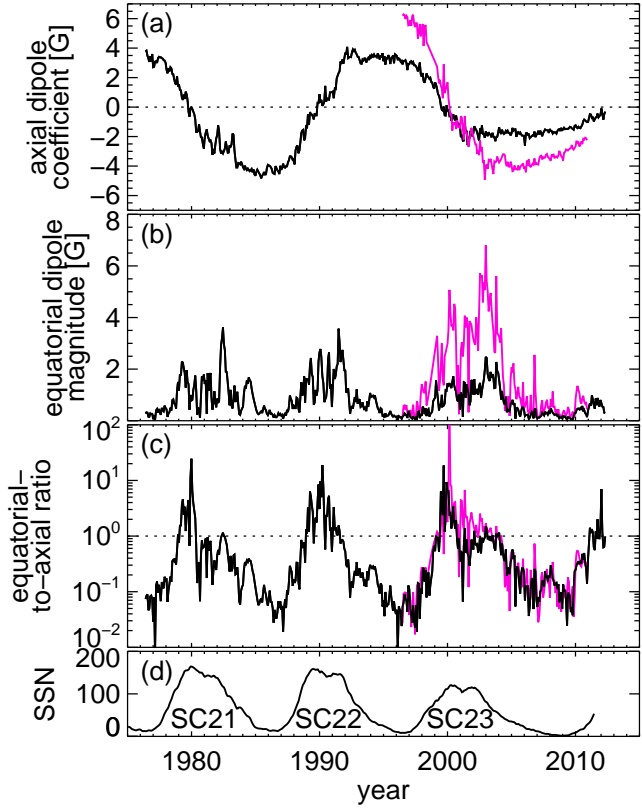}

  \caption{Evolution of the dipole ($\ell=1$) modes, as characterized by the
    (a) axial dipole coefficient $B_1^0$, (b) equatorial dipole magnitude
    $\sqrt{(B_1^{-1})^2+(B_1^1)^2}$, and (c) the ratio of their energies
    $[(B_1^{-1})^2+(B_1^1)^2]/(B_1^0)^2$ for the WSO (black) and MDI
    polar-corrected (magenta) data sets.  Panel (d) shows the monthly smoothed
    sunspot number (SSN) from Solar Influences Data Center at the Royal
    Observatory of Belgium.  The WSO data have not been corrected for known
    saturation effects that reduce the reported values by a factor of 1.8
    \citep{sva1978}.}

  \label{fig:dipole-plots}

\end{figure}

\begin{figure}
  \epsscale{1.0}
  \plotone{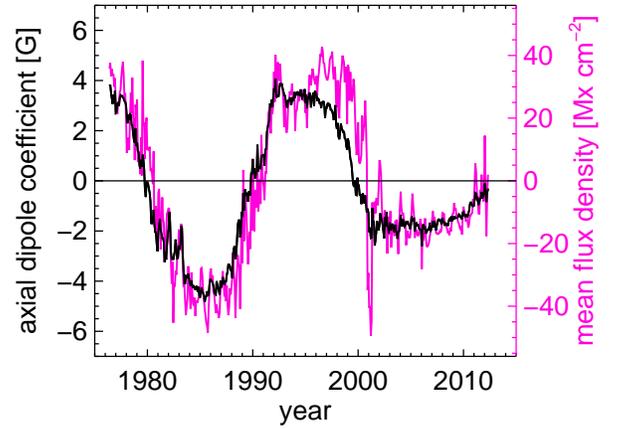}

  \caption{Northern hemispheric mean flux density (magenta) and axial dipole
    coefficient from WSO (black), illustrating the connection between the
    axial dipole and the flux in each hemisphere.  The downward spike in the
    mean hemispheric flux density occurring in 2001 is likely related to WSO
    sensitivity issues occurring during that time period, and may not be
    real.}

  \label{fig:hemflux}

\end{figure}

Because the WSO data span three sunspot activity cycles, a bit of historical
perspective on the evolution of the dipole can be gained, as shown in
Figure~\ref{fig:dipole-plots}.  Figure~\ref{fig:dipole-plots}(a) shows the
amplitude of the axial dipole moment since mid 1976 and its evolution as a
function of the activity level, represented in the figure by the sunspot
number\footnote{Sunspot numbers with slightly different calibrations are
  available from various sources worldwide.  In this article, we use the
  indices provided by the Solar Influences Data Center at the Royal
  Observatory of Belgium, whose sunspot index data are available online at
  \url{http://www.sidc.be/sunspot-index-graphics/sidc\_graphics.php}.} (SSN).
It is also evident that, during the most recent minimum prior to Cycle~24, the
magnitude of the axial dipole component is much lower than during any of the
three previous minima (i.e., those preceding Cycles~21--23).  The connection
between the axial dipole component and the flux in the northern hemisphere is
illustrated in the time-history of the mean flux density as integrated over
the northern hemisphere, shown in Figure~\ref{fig:hemflux}.  We observe a
slight lag for the northern hemisphere averaged magnetic flux with respect to
the axial dipole coefficient due to the contribution of other axisymmetric
modes possessing a different phase.

Figure~\ref{fig:dipole-plots}(b) illustrates the magnitude of the equatorial
dipole since mid 1976.  In step with the relatively lower number of active
regions during Cycle~23 when compared with the maxima for Cycles~21 and~22,
the equatorial dipole strength is found to be lower during the most recent
maximum than during the maxima corresponding to Cycles~21 or~22.

Given the variation in sunspot cycle strengths throughout the past few
centuries, we suspect that cycle-to-cycle variations in the magnitudes of the
axial and equatorial modes are not unusual.  Proxies of the historical
large-scale magnetic field, such as cosmic-ray induced variations of isotopic
abundances measured from ice-core data \citep{ste2012}, also show such
longer-term variation and thus seem to be consistent with this view.
Interestingly, the range over which the variation in the ratio of the energies
possessed by the equatorial versus the axial dipole components is about the
same for the three sunspot cycles observed by WSO, as shown in
Figure~\ref{fig:dipole-plots}(c).  Longer-term measurements of this ratio
unfortunately are not available due to the lack of a sequence of long-term
magnetogram maps to which the harmonic decomposition analysis outlined in
Section~\ref{sec:observations} can be applied.

\subsection{Reversals of the Dipole}

\begin{figure}
  \epsscale{1.0}
  \plotone{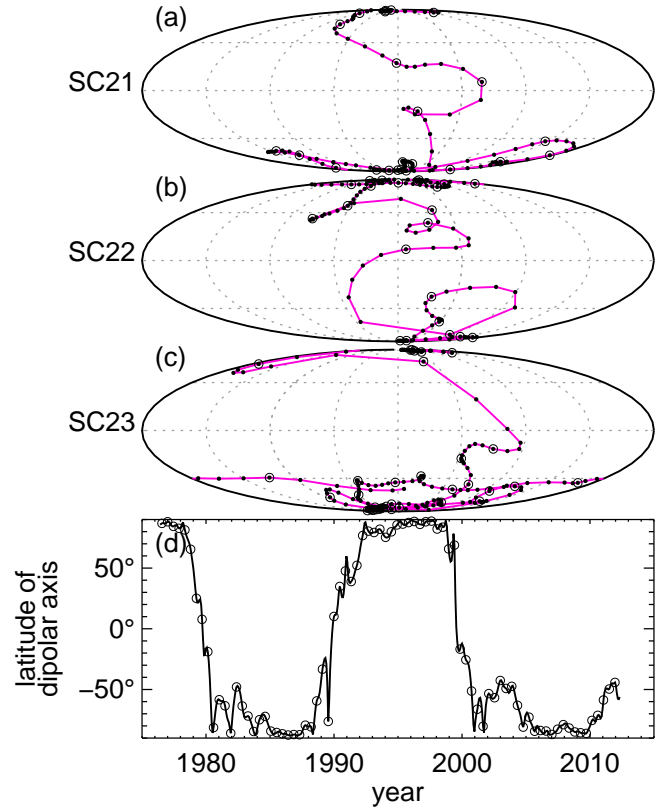}

  \caption{(a)--(c) Mollweide projections of the location of the dipole axis
    for the past three sunspot cycles (Cycles~21--23), as determined from WSO
    synoptic charts.  The solid circles indicate the longitude and latitude of
    the dipole axis for each Carrington rotation, with every sixth Carrington
    rotation also indicated by an open circle.  Grid lines (dashed) are placed
    every 45$^\circ$ in latitude and longitude for reference.  The Carrington
    longitudes of the central meridians of each projection are chosen to best
    illustrate the reversals, and differ in each of the panels.  Panel~(d)
    illustrates the latitude of the dipole axis as a function of time.  The
    open circles in panel~(d) correspond to same times as the open circles in
    panels~(a)--(c).}

  \label{fig:dipole-reversal}

\end{figure}

The process by which old polar caps are canceled out and replaced with new,
opposite-polarity polar caps, as described in the previous section, manifests
itself as a change in sign of the axial dipole amplitude throughout the course
of a sunspot cycle.  Such \textsl{dipole reversals} for the past three sunspot
cycles are shown in Figure~\ref{fig:dipole-reversal}, where the latitude and
longitude of the dipole axis are plotted with time.  It is found that the
dipole axis spends much of its time in the polar regions, and for only about
12--18 months during these cycles is it located equatorward of
$\pm$45$^\circ$.

During these reversals, which occur at or near maximal activity intervals, the
energy in the dipole never completely disappears.  We find that the reduction
in the energy $(B_1^0)^2$ in the the axial dipole is partially offset by an
increase in the energy $(B_1^{-1})^2 + (B_1^1)^2$ in the equatorial dipole.
This results in a reduction of the total energy $\sum_m (B_1^m)^2$ in all
dipolar modes only by about an order of magnitude from its
axial-dipole-dominated value at solar minimum, as shown in
Figure~\ref{fig:di-vs-quad}(a).

Figure~\ref{fig:dipole-reversal} indicates that, during a reversal when the
axial dipolar component is weak, the axis of the equatorial dipolar component
wanders in longitude.  This seemingly aimless wandering occurs because the
longitude of the dipole axis is primarily determined by an interplay amongst
the strongest active regions on the photosphere at the time of observation. As
older active regions decay and newer active regions emerge onto the
photosphere, the equatorial-dipolar axis responds in kind.

\subsection{Quadrupole Field (Modes with $\bs{\ell}=2$)}

\begin{figure}
  \epsscale{1.0}
  \plotone{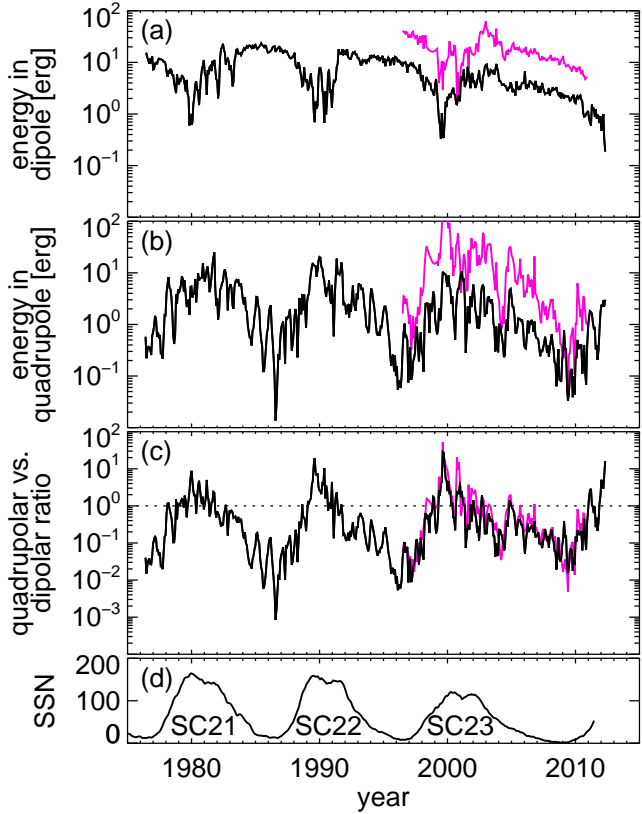}

  \caption{Total energy in (a) dipolar modes $\sum_m (B_1^m)^2$, (b)
    quadrupolar modes $\sum_m (B_2^m)^2$, and (c) their ratio $\sum_m
    (B_2^m)^2/\sum_m (B_1^m)^2$ for the WSO (black) and MDI polar-corrected
    (magenta) data sets.  Panel (d) shows the monthly smoothed SSN.  The WSO
    data have not been corrected for known saturation effects that reduce the
    reported values by a factor of 1.8 \citep{sva1978}.}

  \label{fig:di-vs-quad}

\end{figure}

\begin{figure}
  \epsscale{1.0}
  \plotone{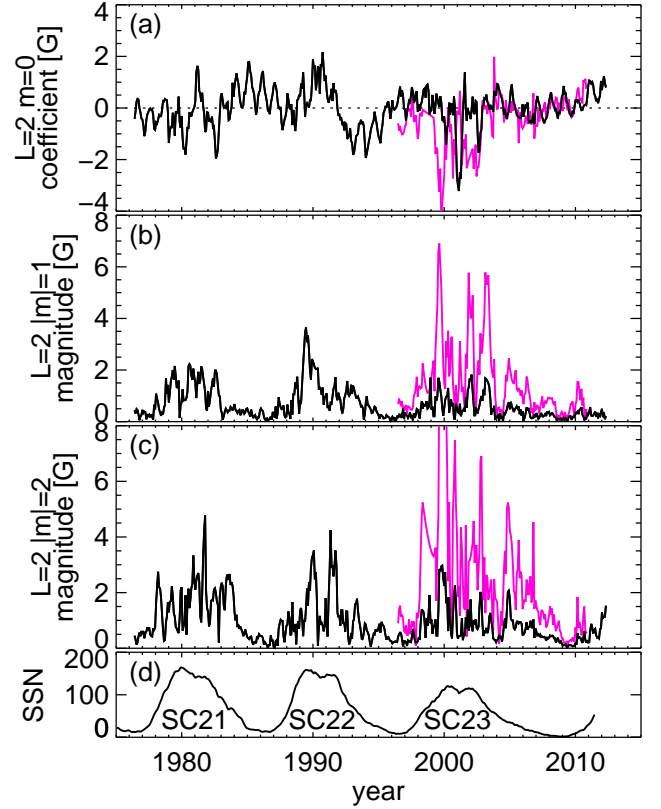}

  \caption{Evolution of the quadrupolar ($\ell=2$) modes, as characterized by
    the (a) axial quadrupole coefficient $B_2^0$, along with higher-order
    magnitudes of the (b) $m=\pm 1$ modes $\sqrt{(B_2^{-1})^2+(B_2^1)^2}$ and
    (c) $m=\pm 2$ modes $\sqrt{(B_2^{-2})^2+(B_2^2)^2}$, for the WSO (black)
    and MDI polar-corrected (magenta) data sets.  Panel (d) shows the monthly
    smoothed SSN.  The WSO data have not been corrected for known saturation
    effects that reduce the reported values by a factor of 1.8
    \citep{sva1978}.}

  \label{fig:quadrupole}

\end{figure}

\begin{figure}
  \epsscale{1.0}
  \plotone{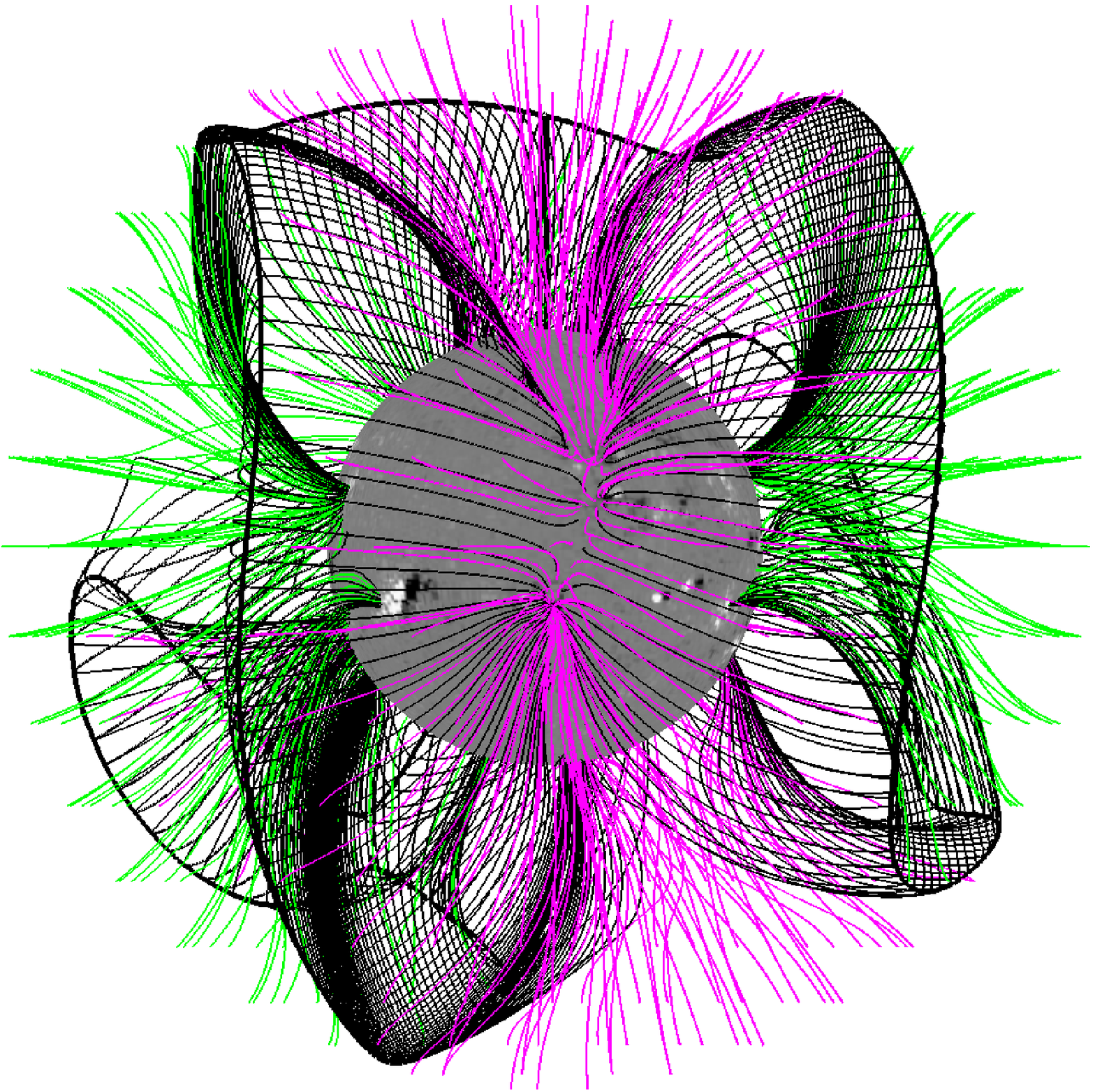}

  \caption{Representation of the coronal magnetic field in October 2000 for
    which the large-scale field is predominantly quadrupolar.  This field is
    the result of a potential-field source-surface extrapolation
    \citep{sch1969} with an upper boundary of 2.5~$R_\odot$ at which the
    coronal field is assumed purely radial.  Both closed (black) and open
    (magenta and green, depending on polarity) field lines are shown in the
    model.  Also shown is the contour of $B_r=0$ at $R=2.5 R_\odot$ (thicker
    black line).}

  \label{fig:quadexample}

\end{figure}

The evolution of the energy contained in the quadrupolar ($\ell=2$) modes
exhibit much more variation than the dipole, as shown in
Figure~\ref{fig:di-vs-quad}. As with the equatorial dipole components, all of
the quadrupolar modes have more power during greater activity intervals than
during quieter periods, as illustrated in the evolution of the various
quadrupolar modes plotted in Figure~\ref{fig:quadrupole}.  Furthermore, when
large amounts of activity occur, it is possible for the total energy $\sum_m
(B_2^m)^2$ in all quadrupolar modes to be greater than the energy $\sum_m
(B_1^m)^2$ in the dipolar modes at the photosphere.  The ratio between these
two groups of modes is shown in Figure~\ref{fig:di-vs-quad}(c), from which it
is evident that during each of the past three sunspot cycles there have been
periods of time when the quadrupolar energy exceeded the dipolar energy by as
much as a factor of 10.  The corona, in turn, reflects the relative strength
of a strong quadrupolar configuration of photospheric magnetic fields by
creating complex sectors and possibly multiple current sheets that extend into
the heliosphere.  One example of such complex field geometry is suggested by
the potential-field source-surface model of Figure~\ref{fig:quadexample},
where a quadrupolar configuration having an axis of symmetry lying almost in
the equatorial plane is seen to predominate.


\subsection{Octupole Field (Modes with $\bs{\ell}=3$)} \label{sec:octupole}

\begin{figure}
  \epsscale{1.0}
  \plotone{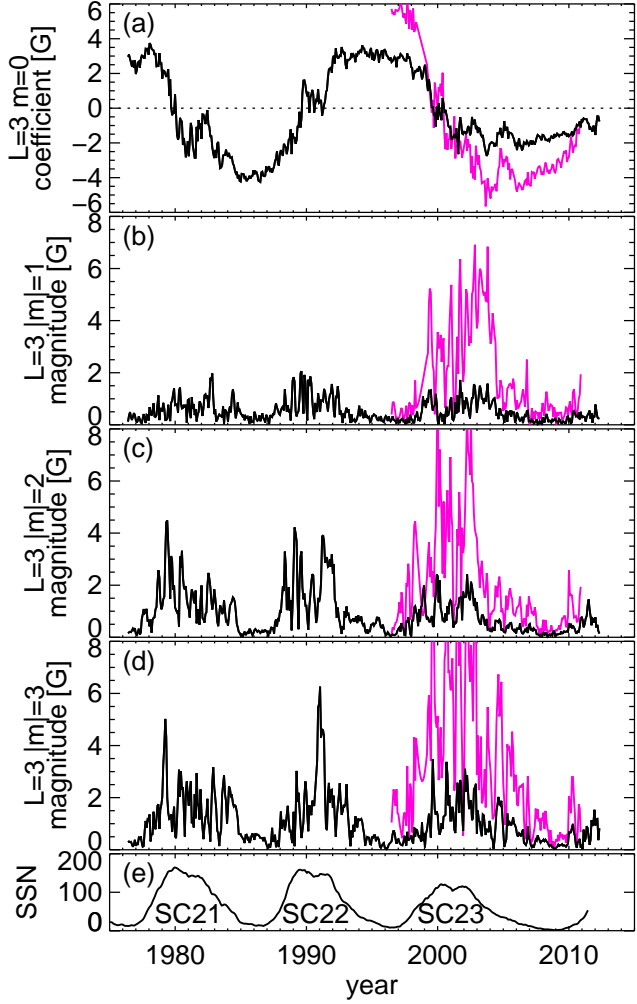}

  \caption{Evolution of the octupolar ($\ell=3$) modes, as characterized by
    the (a) axial octupole coefficient $B_3^0$, along with higher-order
    magnitudes of the (b) $m=\pm 1$ modes $\sqrt{(B_3^{-1})^2+(B_3^1)^2}$, (c)
    $m=\pm 2$ modes $\sqrt{(B_3^{-2})^2+(B_3^2)^2}$, and (d) $m=\pm 3$ modes
    $\sqrt{(B_3^{-3})^2+(B_3^3)^2}$, for the WSO (black) and MDI
    polar-corrected (magenta) data sets.  Panel (e) shows the monthly smoothed
    SSN.  The WSO data have not been corrected for known saturation effects
    that reduce the reported values by a factor of 1.8 \citep{sva1978}.}

  \label{fig:octupole}

\end{figure}

As with the quadrupole, the octupolar modes contain more power during periods
of high activity and less power during minimum conditions, as illustrated in
Figure~\ref{fig:octupole}.  The exception is the axial octupolar coefficient
$B_3^0$, plotted in panel~(a) of Figure~\ref{fig:octupole}, which is nonzero
during solar minima and exhibits sign reversals during sunspot maxima in a
manner similar to the axial dipole coefficient $B_1^0$.

The behavior of the various $m=0$ modes can be understood by considering their
functional symmetry: the $Y_\ell^0$ functions are \textsl{antisymmetric} in
$\theta$ (i.e., antisymmetric across the equator) when the degree $\ell$ is
odd, whereas for even $\ell$ the $Y_\ell^0$ functions are \textsl{symmetric}
in $\theta$.  The presence of polar caps during solar minimum, a highly
antisymmetric configuration, is reflected in the similar evolution of the
$B_1^0$ and $B_3^0$ coefficients, which correspond to the axial dipole
($\ell=1$, $m=0$) and octupole ($\ell=3$, $m=0$) modes.  The axial quadrupole
($\ell=2$, $m=0$) mode does not share this behavior because, as a symmetric
mode, it is not sensitive to the presence of the polar caps during solar
minima.

\begin{figure}
  \epsscale{1.0}
  \plotone{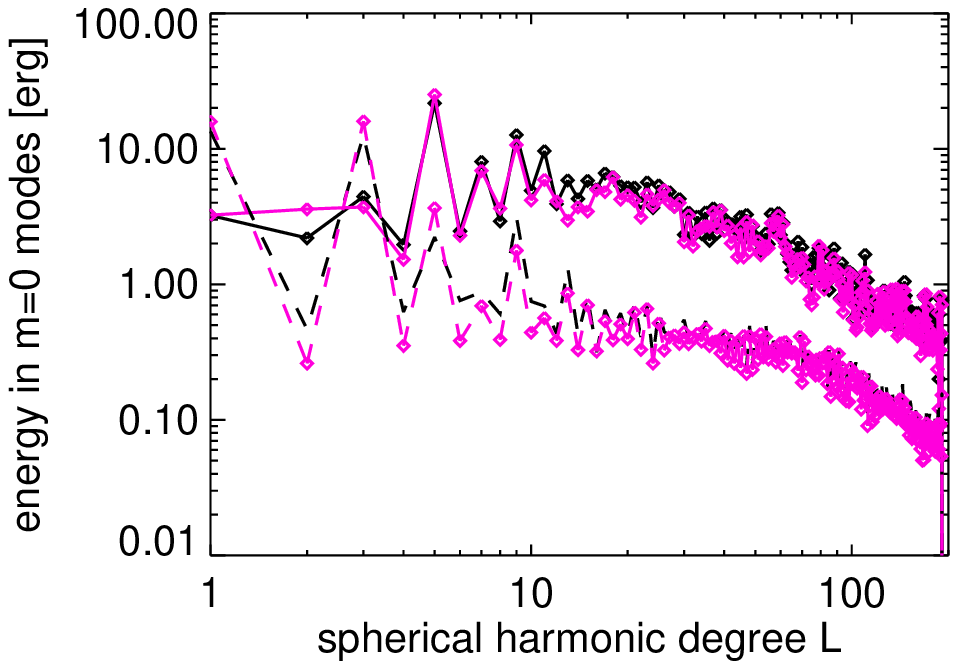}

  \caption{Time-averaged energies in the axisymmetric modes
    $\overline{(B_\ell^0)^2}$ as a function of $\ell$ for MDI original (black)
    and polar-corrected (magenta) data sets, for more active conditions (solid
    lines; defined as when SSN>100) and for quieter periods (dashed lines;
    defined as when SSN<50).  The interpolation scheme used to correct the MDI
    polar flux is described in \citet{sun2011}.}

  \label{fig:energyspectrum-axi}

\end{figure}

The dependence of the $B_\ell^0$ coefficients on the degree $\ell$ is
illustrated in Figure~\ref{fig:energyspectrum-axi}, where the time-averaged
energies from the MDI data (spanning Solar Cycle 23) as a function of degree
$\ell$ are plotted.  Prior to averaging, the spectra were placed in two
classes: Carrington rotations for which the SSN is relatively large (defined
as when SSN>100) and rotations for which the SSN is relatively small (defined
as when SSN<50), thus capturing the state of the Sun when it is either overtly
active or overtly quiet.  The figure indicates that the even-odd behavior is
more pronounced during quiet periods, and these occur near and during solar
minimum when the polar-cap field is significant.  During active periods the
even-odd trend is still recognizable, but because the polar caps are weak and
the active regions are primarily oriented east-west (i.e., in the equatorial
plane and thus contributing little power to the axisymmetric modes) the
even-odd trend is less pronounced.  We will further discuss the behavior of
axisymmetric modes in the context of Babcock-Leighton dynamo models in
Section~\ref{sec:bl-axisymmetric}.

\begin{figure}
  \epsscale{1.0}
  \plotone{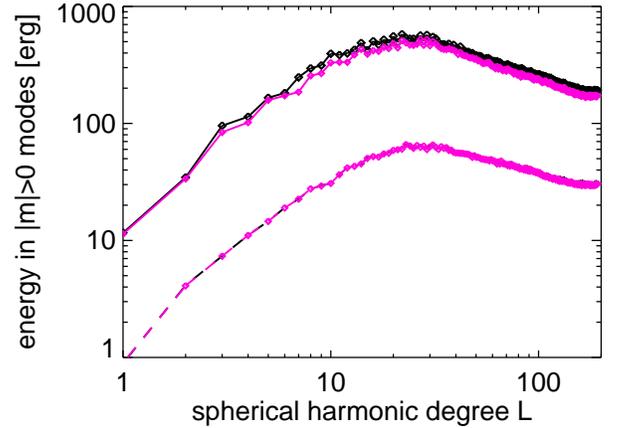}

  \caption{Time-averaged energies in the non-axisymmetric modes
    $\overline{\sum_{m>0}(B_\ell^m)^2}$ as a function of $\ell$ for MDI
    original (black) and polar-corrected (magenta) data sets, for more active
    conditions (solid lines; defined as when SSN>100) and for quieter periods
    (dashed lines; defined as when SSN<50).  The interpolation scheme used to
    correct the MDI polar flux is described in \citet{sun2011}.}

  \label{fig:energyspectrum-nonaxi}

\end{figure}

\subsection{Full Spectra and Most Energetic Modes}

One property of the spherical harmonic functions $Y_\ell^m(\theta,\phi)$ is
that the degree $\ell$ is equal to the number of node lines (i.e., contours in
$\theta$ and $\phi$ where $Y_\ell^m=0$). In other words, the spatial scale
represented by any harmonic mode (i.e., the distance between neighboring node
lines) is determined by its spherical harmonic degree $\ell$. As a result, the
range of $\ell$ values containing the greatest amount of energy indicates the
dominant spatial scales of the magnetic field.

To this end, we have averaged the non-axisymmetric power spectra from each of
the datasets both over time and over $m$, and have displayed the result in
Figure~\ref{fig:energyspectrum-nonaxi}.  As with
Figure~\ref{fig:energyspectrum-axi}, we have divided the spectra into active
and quiet classes depending on SSN.  In the figure, it can be seen that the
magnetic power spectra form a broad peak with a maximum degree occurring at
$\ell_{P_\text{max}}\approx$25, corresponding to a size scale of about
360$^\circ/\ell_{P_\text{max}}\approx$15$^\circ$ in heliographic coordinates.
Stated another way, this indicates that much of the magnetic energy can be
found (not surprisingly) on the spatial scales of solar active regions or
their decay products.

Energy spectra determined from WSO charts (not shown) do not show the same
broad peak at $\ell_{P_\text{max}}\approx$25 as found in the curves from the
MDI-derived data shown in Figure~\ref{fig:energyspectrum-nonaxi}.  This is an
effect of the lower spatial resolution of the WSO magnetograph (which has
180$\arcsec$ pixels, and is stepped by 90$\arcsec$ in the east-west direction
and 180$\arcsec$ in the north-south direction when constructing a magnetogram)
versus MDI (which has a plate scale of 2$\arcsec$ per pixel in full-disk
mode).  The WSO magnetograph, as a result, does not adequately resolve modes
higher than about $\ell=15$, creating aliasing effects even at moderate values
of $\ell$ in the energy spectra.  Accordingly, as longer time series of data
from newer, higher-resolution magnetograph instrumentation are assembled, the
high-$\ell$ behavior of the energy spectra (such as those shown in
Figs.~\ref{fig:energyspectrum-axi} and~\ref{fig:energyspectrum-nonaxi}) may
change due to better observations of finer scales of magnetic field.

\subsection{Primary and Secondary Families} \label{sec:families}

The projection of the solar surface magnetic fields onto spherical harmonic
degrees allows us to delineate the main symmetries of the magnetic field.  As
noted in \S\ref{sec:observations}, the harmonic modes can be classified as
either \textsl{axisymmetric} ($m=0$) or \textsl{non-axisymmetric} ($m \neq
0$).  Separately, the harmonic modes can be either \textsl{antisymmetric} (odd
$\ell + m$) or \textsl{symmetric} (even $\ell + m$) with respect to the
equator \citep{gub1993}.  Some authors refer to antisymmetric modes as
\textsl{dipolar} and symmetric modes as \textsl{quadrupolar} (presumably
because the axial dipole and quadrupole modes usually possess the most power
in their respective categories), while others synonymously assign these modes
to either the \textsl{primary} and \textsl{secondary} family (e.g.,
\citealt{mcf1991} when characterizing the Earth's magnetic field geometry),
respectively.  In this article, we adopt the primary- and secondary-family
nomenclature when describing the equatorial symmetry because this avoids the
confusion that may otherwise occur when, for example, it is realized that the
equatorial dipole mode ($\ell=1, m=1$) formally belongs to the ``quadrupolar''
family of modes (since $\ell + m$ is even for this mode).

One important result put forward by the geomagnetic community is that the
relative strengths of the primary and secondary families are different during
geomagnetic field reversals and excursions.  During reversals, the modes
associated with the secondary family predominate over primary-family modes,
and during excursions this is not the case \citep{leo2007}.  We now
investigate whether analogous behavior is occurring in the solar setting, by
determining which harmonic modes are most correlated with the axial dipole and
axial quadrupole.


When applied to two variables, the Spearman rank correlation index
$\rho\in[-1,1]$ indicates the degree to which two variables are monotonically
related.  The index $\rho$ is positive when both variables tend to increase
and decrease at the same points in time.  A rank correlation analysis is more
general than a more common Pearson correlation analysis, which specifically
measures how well two variables are linearly related, whereas the rank
correlation analysis enables a determination of whether the time evolution of
two mode amplitudes follow a similar pattern in time without regard to their
(unknown) functional dependency.

In Table~\ref{table:rankmatrix} we list the degrees $\ell$ and orders $m$
corresponding to the harmonic coefficients $B_\ell^m(t)$ that have the highest
$\rho$ (positive correlation) when compared with the axial dipole and axial
quadrupole coefficients $B_1^0(t)$ and $B_2^0(t)$ (which peak at different
phases of the sunspot cycle).  The corresponding harmonic modes comprise the
strongest modes in the primary and secondary families, respectively.  We find
that, among the mode amplitudes that are positively correlated with $B_1^0$,
two out of three belong to the primary family.  Similarly, for $B_2^0$, 7 of
the 10 most-correlated modes are members of the secondary family.

\begin{figure}
  \epsscale{1.0}
  \plotone{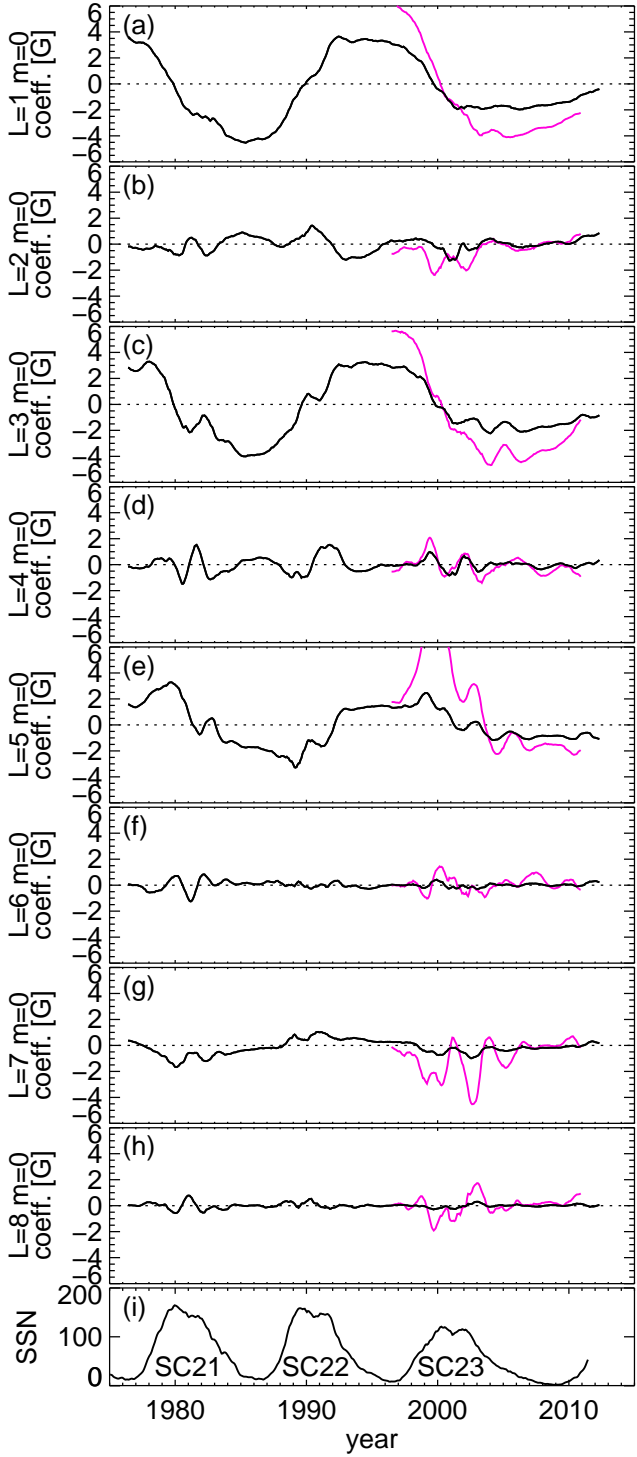}

  \caption{(a)--(h) Coefficients $B_\ell^0$ for the axisymmetric modes of the
    first eight degrees $\ell$ as a function of time, as calculated from the
    WSO (black) and MDI polar-corrected (magenta) synoptic maps and after
    boxcar smoothing with a width of 1 yr.  Panel (i) shows the monthly
    smoothed SSN.  The WSO data have not been corrected for known saturation
    effects that reduce the reported values by a factor of 1.8
    \citep{sva1978}.}

  \label{fig:axisymmetric}

\end{figure}

\begin{figure}
  \epsscale{0.9}
  \plotone{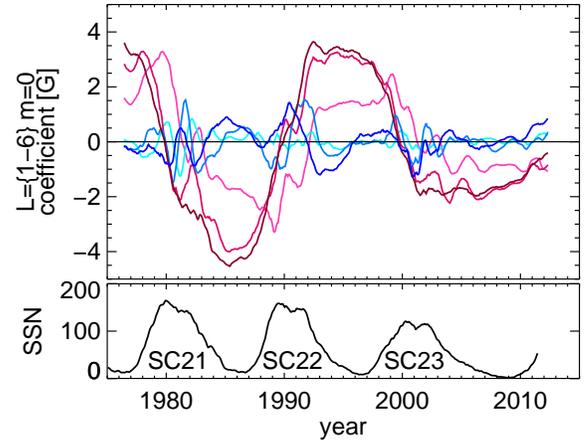}

  \caption{Overplotted coefficients $B_\ell^0$ from
    Figure~\ref{fig:axisymmetric} of the first 3 odd ($\ell=\{1,3,5\}$; dark
    red, red, light red lines, respectively) and even ($\ell=\{2,4,6\}$; dark
    blue, blue, light blue, respectively) axisymmetric modes, as calculated
    from WSO synoptic maps.}

  \label{fig:axisymmetric-overplot}

\end{figure}

These correlations indicate a preference in the solar dynamo, at least as
inferred from its surface characteristics, for modes belonging to the same
family and thus having the same north-south symmetry characteristics to be
excited nearly in phase.  This preference is demonstrated further in
Figures~\ref{fig:axisymmetric} and~\ref{fig:axisymmetric-overplot}, in which
the long-term trends of the time evolution of the first several axisymmetric
mode coefficients are shown, after smoothing with a boxcar filter having a
width of 1~yr.  (We focus here on the axisymmetric mode properties because
these modes are the only ones considered in most mean-field dynamo models, as
discussed further in \S\ref{sec:bl-axisymmetric}.)  In
Figure~\ref{fig:axisymmetric}, there is a clear correlation amongst the first
few odd-$\ell$ and amongst the first couple of even-$\ell$ mode coefficients,
a trend which is emphasized in Figure~\ref{fig:axisymmetric-overplot} in which
these same mode coefficients are overplotted.  We note that the mode groupings
are not precisely in phase, as evidenced for example by the lag in $\ell=3$
and especially the $\ell=5$ modes reversing signs with respect to the $\ell=1$
mode.  When $\ell\ge 6$ or so, these trends become much weaker amongst the
axisymmetric modes (although Table~\ref{table:rankmatrix} indicates that this
is not necessarily true for the non-axisymmetric modes), presumably because as
smaller and smalle1r scales are considered the effects of the global
organization associated with the 11-year sunspot cycle are less important in
structuring the surface magnetic field.

Figure~\ref{fig:axisymmetric-overplot} additionally illustrates that the modes
of the secondary family attain amplitudes of about 25\% of the primary mode
amplitudes.  Furthermore, the primary and secondary mode families are out of
phase: during reversals the primary modes become weak at the same time as the
amplitudes of the modes associated with the secondary family become maximal,
which was shown previously in Figure~\ref{fig:di-vs-quad}(c).  This same
pattern is observed to occur during reversals of the axial dipole field of the
geodynamo.  As in the geodynamo case, we ascribe the relative amplitudes and
phase relation between the primary and secondary families observed during
solar dipole reversals as a strong indication that the interplay of the mode
families play a key role in the process by which the axial dipole reverses.
Hence, any realistic model of the solar dynamo must excite both families of
modes to similar amplitude levels, and must exhibit similar coupling between
modes belonging to the primary and secondary families.

\begin{table}[!ht]
\begin{center}
Most Correlated modes\\
\begin{tabular}{cccccc}
\hline
\hline
& $\ell=1$, $m=0$ & & & $\ell=2$, $m=0$ & \\
\hline
$\ell$ & $m$ & Primary & $\ell$ & $m$ & Secondary \\
\hline
\hline
3 & 0 & Y & 4 & 0 & Y \\
5 & 0 & Y & 4 & 1 & N \\
2 & 0 & N & 9 & 9 & Y \\
  &   &   & 6 & 1 & N \\
  &   &   & 7 & 0 & N \\
  &   &   & 9 & 1 & Y \\
  &   &   & 5 & 1 & Y \\
  &   &   & 1 & 1 & Y \\
  &   &   & 7 & 7 & Y \\
  &   &   & 3 & 3 & Y \\
\hline
\hline
\end{tabular}
\end{center}
\caption[dominant modes]{\label{dom_mod} Ranking of the most positively
  correlated modes within the primary and secondary families for which $\ell
  \le 9$.  The basis for comparison in each family is the lowest-degree
  axisymmetric mode belonging to each family, specifically the axial dipole
  ($\ell=1$, $m=0$) and axial quadrupole ($\ell=2$, $m=0$) modes for the
  primary and secondary families, respectively.  The most correlated mode is
  the next axisymmetric mode in each family.  The equatorial dipole mode
  ($\ell=1$, $m=1$) is more correlated with the axisymmetric quadrupole, as
  expected from its symmetry properties.  We note the presence of 4 sectoral
  modes (for which $\ell=m$) in the list of the secondary family.}
\label{table:rankmatrix}
\end{table}

\section{Theoretical Implications for Solar Dynamo} \label{sec:discussions}

As demonstrated in previous sections, the amplitudes of the various harmonic
modes of the solar magnetic field are continually changing.  During reversals,
as the axial dipole necessarily undergoes a change in sign, other modes
predominate such that the amplitude of the solar magnetic field never vanishes
during a reversal.  As a result of such reversals occurring in the middle of
each 11-yr sunspot cycle, during the rising phase of each cycle the polar
fields and the emergent poloidal fields have opposite polarity
\citep{bab1961,ben2004}, while in the declining phase the polarity of sunspots
and active regions are aligned with the newly formed polar-cap field.

We have already noted how the temporal modulation of the large-scale harmonic
modes comprising the primary and secondary families during polarity reversals
appears similar to that of the magnetic field of the Earth
\citep{mcf1991,leo2007}.  We have illustrated that as the magnitude of the
primary-family mode amplitudes (primarily those of the axisymmetric odd-$\ell$
modes $B_\ell^0$) lessen, the secondary-family mode amplitudes (particularly
those of the equatorial dipole $B_1^1$ and axial quadrupole $B_2^0$)
simultaneously increase.  Once the secondary-family modes have peaked, the
primary-family modes grow as a result of the growing polar-caps.  Such
interplay between primary and secondary families provides insight toward an
understanding of the processes at play in the solar dynamo that are assumed to
be responsible for the occurrence of the observed cyclic activity
\citep{tob2002}.

The presence of power in members of both the primary and secondary families
indicates that the solar dynamo excites modes that are both symmetric and
antisymmetric with respect to the equator.  As was demonstrated by
\citet{rob1972}, this cannot occur unless nonlinearities exist or unless basic
ingredients of the solar dynamo (such as, for example the $\alpha$ and/or
$\omega$ effects, or the meridional flow) possess some degree of north-south
asymmetry.  In light of the the parameter regime in which the solar dynamo is
thought to operate, including large fluid and magnetic Reynolds numbers $R_e$
and $R_m$ believed to characterize the solar convection zone (both of order
$10^{12}$--$10^{15}$; \citealt{sti2002,oss2003}), one expects the Sun to
possess a nonlinear dynamo.  Detailed observations of the magnetic field in
the solar interior where dynamo activity is thought to occur are not
available, but the observed magnetic patterns and evolution provide
circumstantial evidence of turbulent, highly nonlinear processes that lead to
complex local and nonlocal cascades of energy and magnetic helicity
\citep{ale2005,ale2007,liv2010,str2012}.  Yet, the presence of regular
patterns formed by the emergent flux on the solar photosphere, as codified by
Hale's Polarity Law, Joy's Law for active-region tilts, and the approximately
regular cycle lengths, suggest that some ordering is indeed happening in the
solar interior.

With the aim of distilling the necessary elements of the various nonlinear
dynamos into a manageable framework, multiple authors have created idealized
models of the solar dynamo, including (for example)
\citet{wei1984,fey1990,ruz1992,kno1996,tob1997,kno1998,mel2001,wei2000,spi2009}.
Similarly, for the geodynamo there are many efforts, including
\citet{gla1995,hei2005,chr2006,bus2008,nis2008,tak2008,chr2010}.  A completely
different approach has been taken by \citet{pet2008a} and \citet{pet2009}, who
have developed simplified models of the von Karman sodium (VKS) laboratory
experiment \citep{mon2007}.  In all of these idealized models, the modulations
resulting from the coupling between magnetic modes from the different
families, or between the magnetic field and fluid motions \citep{tob2002} can
be analyzed in terms of the equations that describe the underlying systems.
The variability of the most prominent cycle period develops as a result of the
coupling of modes introducing a second time scale into the dynamo system,
often leading to a quasi-periodic or chaotic behavior of the magnetic field,
cycle length, and/or dominant parity.  One can further understand via symmetry
considerations how reversals and excursions arise \citep{gub1993}.

\subsection{Reversals and Coupling between Modes}

\begin{figure}
  \epsscale{0.9}
  \plotone{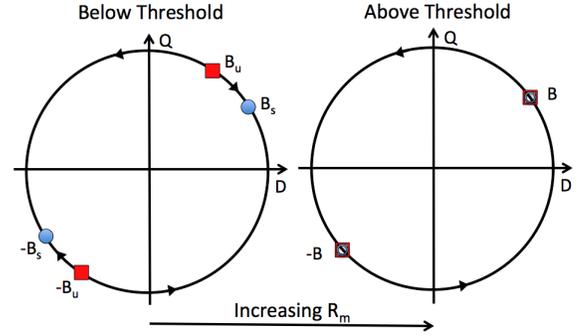}

  \caption{Schematic diagrams (adapted from \citealt{pet2009}) of a magnetic
    dynamo system on either side of a saddle-node bifurcation, with two
    distinct polarity configurations represented by $B$ and $-B$.  The
    coordinate axes represent states where the primary (as represented by the
    axial dipole $D$) or secondary (as represented by the axial quadrupole
    $Q$) families are dominant.  In the left-hand diagram, stable ($\pm B_s$)
    and unstable ($\pm B_u$) states present during the system's evolution are
    indicated by blue circles and red squares.  Perturbations away from a
    stable point can either cause the system to evolve to the opposite stable
    configuration (if the perturbation is strong enough) or simply cause an
    excursion in which the system returns to the same stable state.  In the
    right-hand diagram, corresponding to the same system at a higher magnetic
    Reynolds number $R_m$ the stable and unstable points have merged, and the
    system simply oscillates between the two configurations in a limit cycle.}

  \label{fig:bifurcation-diagram}

\end{figure}

To illustrate such a dynamical system, following the work of \cite{pet2008a},
we assume that the axial dipole and quadrupole modes are nonlinearly coupled.
We can then define a variable $A(t) = B_1^0(t) + i B_2^0(t)$, where we have
used the time-varying mode coefficients defined in
Equation~\ref{eq:expansion}, and write an evolution equation that satisfies
the symmetry invariance found in the induction equation, i.e., $\bs{B}
\rightarrow -\bs{B}$.  It then follows that the symmetry $A \rightarrow -A$
must also be satisfied, and that such a equation to leading order is
\begin{equation}
\frac{dA}{dt} = \mu_1 A + \mu_2 \bar{A} + \nu_1 A^3 + \nu_2 A^2 \bar{A} +
\nu_3 A \bar{A}^2 + \nu_4 \bar{A}^3,
\end{equation}
where $\mu_i$ and $\nu_i$ are complex coefficients and $\bar{A}=B_1^0 - i
B_2^0$ is the complex conjugate of $A$, and the quadratic terms have vanished
due to symmetry considerations.  As discussed in \citet{pet2008a} and
\citet{pet2009}, such dynamical systems are subject to bifurcations.  In
particular, they demonstrate that this dynamical system can be characterized
by a saddle-node bifurcation when comparing its properties with so-called
normal form equations \citep{guc1982}.  In such a bifurcated system, both
stable and unstable equilibria (fixed points) exist, as illustrated in the
left panel of Figure \ref{fig:bifurcation-diagram}.  For instance, if the
solution lies at a stable point (for example, where the dipole axis is
oriented northward), fluctuations in the system may disturb the equilibrium
and push the magnetic axis away from its stable location.  If such
fluctuations are not strong enough, the evolution of the dynamical system
resists the deterministic evolution of the system and the system returns to
its original configuration (in the example, resulting in an excursion of the
dipole), such as seen in the geomagnetic field \citep{leo2007}.  If instead
the fluctuations are large enough to push the system past the unstable point,
the magnetic field then evolves toward the opposite stable fixed point allowed
by $\bs{B} \rightarrow -\bs{B}$ (in the example, resulting in a reversal that
changes the dipole axis to a southward orientation).  Such behavior is also
seen in the VKS experiment, from which is observed irregular magnetic activity
combined with both excursions and reversals.  Reversals result in an
asymmetric temporal profile, with the dipole evolving slowly away from its
equilibrium followed by a swift flip (cf., Fig.~3 of \citealt{pet2009}).

\begin{figure}
  \epsscale{1.0}
  \plotone{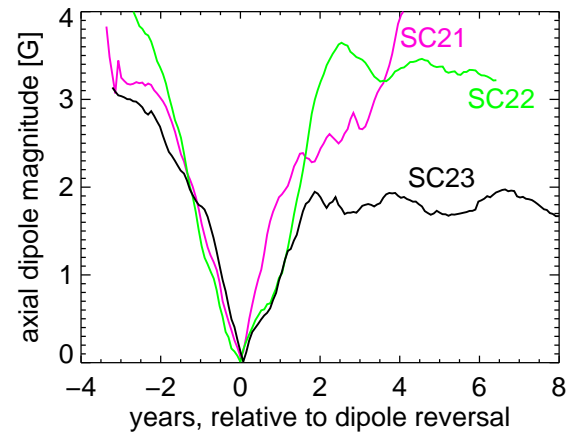}

  \caption{The reversals, as defined by the magnitude of the axial dipole
    component for WSO, for the past sunspot Cycles 21--23.  The reversal for
    these three cycles occurred in Oct.~1979, Nov.~1989, and Jul.~1999,
    respectively.}

  \label{fig:scaled-dipole-reversals}

\end{figure}

In Figure~\ref{fig:scaled-dipole-reversals} we have overplotted the last three
sunspot cycle reversals such that the zero-crossings of the axial dipole
coefficients $B_1^0$ for each cycle are aligned.  It has been recently shown
that the 10 major geomagnetic reversals for which detailed records exist
occurring during the past 180~Myr possess a characteristic shape upon suitable
normalization \citep{val2012}.  This shape can be described as comprising a
precursory event lasting of order 2500~yr, a quick reversal not exceeding
1000~yr, and a rebound event of order 2500~yr.  \citet{pet2009} show that the
magnetic field in a simplified VKS laboratory experiment exhibits differing
behavior during reversals and excursions.  During reversals, the magnetic
field has an asymmetric profile that contains a slow decrease in the dipole,
followed by a rapid change of polarity and buildup of the opposite polarity,
whereas excursions are more symmetric.  Additionally, after reversals the
magnetic dipole overshoots its eventual value before settling down, whereas
during excursions no such overshooting is measured (see Fig.~3 of
\citealt{pet2009}).  For the solar cases displayed in
Figure~\ref{fig:scaled-dipole-reversals}, we find that only the (green) curve
of the reversal of cycle~22 exhibits an overshoot, whereas the other two
cycles do not.  Further, the rates at which the solar dynamo approaches and
recovers from the reversal appear to be equal, leading to a symmetric profile,
in contrast with the VKS results.  Therefore, the Sun seems to reverse its
magnetic field in a less systematic way than other systems that have shown
such behavior.

Analyzing such systems from a dynamical systems perspective, when changing the
control parameter (here, $R_m$) past the bifurcation point, the stable and
unstable points coalesce and merge and the saddle nodes disappear, as shown in
the right panel of Figure~\ref{fig:bifurcation-diagram}.  This act transforms
the system from one containing fixed points to one containing limit cycles
with no equilibria (e.g., \citealt{guc1982}), yielding an oscillatory solution
that manifests itself as cyclic magnetic activity.  Typically, large
fluctuations are required in order to put the dynamical system above the
saddle-node bifurcation threshold.

In the case of the Sun, both the primary and secondary families are excited
efficiently and a strong coupling between them is exhibited.  The model of
\citet{pet2008a}, in spirit very close to the studies of \citet{kno1996} or
\citet{mel2001}, may be used to guide our interpretation of the solar data.
As illustrated in Figure~\ref{fig:axisymmetric-overplot}, the axisymmetric
dipole and quadrupole are out of phase, such that their coupling may lead to
global reversals of the solar poloidal field.  To the best of our knowledge,
however, the solar dynamo does not exhibit excursions of its magnetic field
(unlike the geodynamo) but instead undergoes fairly regular reversals that
take about one or two years to transpire (cf.,
Fig.~\ref{fig:dipole-reversal}). The solar dynamo is thus better approximated
by a model in which a limit cycle is present.  One may presume that the
difference between the geodynamo and the solar dynamo may be a result of the
large degree of turbulence present in the solar convection zone, whereas the
mantle of the Earth has a more laminar convective flow and thus is below the
bifurcation threshold where fixed points are still present.

It may be the case that the solar dynamo is better described by a Hopf
bifurcation, in which a limit cycle arises (branches from a fixed point) as
the bifurcation parameter is changed.  The dynamo instability that occurs as a
result of the interaction of magnetic fields and fluid flows (such as
$\alpha\omega$ dynamos typically used to model the Sun, as summarized in
\citealt{tob2002}) often arises from a Hopf bifurcation.  This allows the
system to pass through domains having different properties, such as the
aperiodic oscillations that characterize the grand minima and nonuniform
sunspot cycle strengths of the solar dynamo (e.g., \citealt{spi2009} and
references therein).  The data analysis shown here does not favor a particular
type of bifurcation, but does indicate efficient coupling between the primary
and secondary families.

Yet another approach toward investigating magnetic reversals is to develop
detailed numerical simulations solving the full set of MHD equations.  Such
three-dimensional numerical simulations in spherical geometry of the Earth's
geodynamo \citep{gla1995,li2002,nis2008,ols2011} or of the solar global dynamo
\citep{bru2004,bro2006,rac2011} have looked at the behavior of the polar
dipole vs multipolar modes.  Even though such models have large numerical
resolution and thus possess a large number of modes, all have the property
that the dominant polarity of the magnetic field follows the temporal
evolution of a few low-order modes, even if in some cases the magnetic energy
spectrum peaks at higher angular degree $\ell$.  These findings suggest that
the coupling between the primary and secondary family remains an important
factor in characterizing polarity reversals for these simulations, and is
thought to be linked to a symmetry-breaking of the convective flow
\citep{nis2008,ols2011}. Some studies of the geomagnetic field (e.g.,
\citealt{cle2004}) even advocate for a coupling between two modes of the same
primary family, such as the axial dipole and octupole.  While in the solar
data these modes are well correlated, the coupling between the primary and
secondary families of modes seems more likely to be at the origin of the
reversal, as demonstrated in \S\ref{sec:families}.

\subsection{Mean-Field Dynamo Models and the Axisymmetric Modes}  \label{sec:bl-axisymmetric}

Mean-field dynamo models are found to capture the essence of the large scale
solar dynamo \citep{mof1978,oss2003,dik2004,cha2010}.  At present, the most
favored model is the mean-field Babcock-Leighton (BL) dynamo model (e.g.,
\citealt{dik2004}), in which the mean magnetic induction equation is solved
using empirical guidance for both the differential rotation and meridional
circulation profiles, as well as for parameterizations of the $\alpha$-effect
and poloidal-field source terms.  In this section, we use the Stellar Elements
(STELEM; T.~Emonet \& P.~Charbonneau, 1998, priv.\ comm.) code (see Appendix~A
of \citealt{jou2007} for more details) to solve the axisymmetric BL dynamo
equations, and investigate some of the consequences of the coupling between
modes from the primary and secondary families. In the interest of brevity, we
refer interested readers to Appendix~\ref{sec:bleqns} for a listing of the
governing equations associated with BL dynamos.

In many BL solar dynamo models, the parameters governing the imposed flows and
the poloidal-field source terms are chosen based on their solar counterparts.
When carefully chosen these terms favor modes from the primary family and thus
antisymmetric field configurations, since this is what the Sun apparently
favors much of the time.  This is a result of the commonly used latitudinal
profiles of the key dynamo ingredients (symmetric large-scale flows and
antisymmetric alpha effect) combined with the parity in the BL mean-field
dynamo equations, leading to a situation where modes of the primary family
remain uncoupled to modes of the secondary family that allows both dynamo
families to coexist without much interaction.  We consider the symmetry of the
BL equations used here in Appendix~\ref{sec:symmetry} (see also
\citealt{rob1972} and \citealt{gub1993} for broader discussions on this
topic).

\begin{figure}
  \epsscale{1.0}
  \plotone{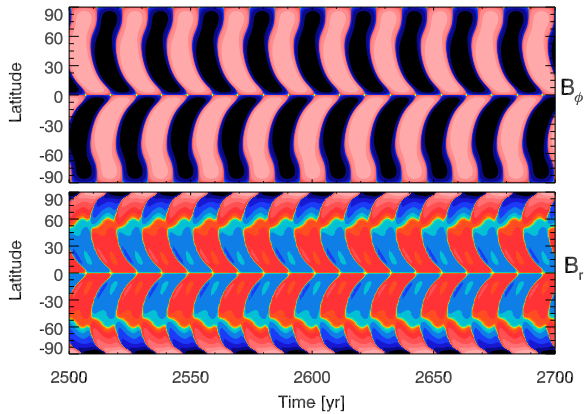}

  \caption{Latitude-time plots of $B_\phi$ (at the tachocline) and $B_r$ (at
    the surface) produced by a mean-field BL dynamo model that uses
    empirical guidance for the solar differential rotation and meridional flow
    profiles, and that is initialized with a dipolar magnetic field.  The
    lower panel is analogous to the standard solar butterfly diagram.}

  \label{fig:bl-dipole-butterfly}

\end{figure}

\begin{figure}
  \epsscale{1.0}
  \plotone{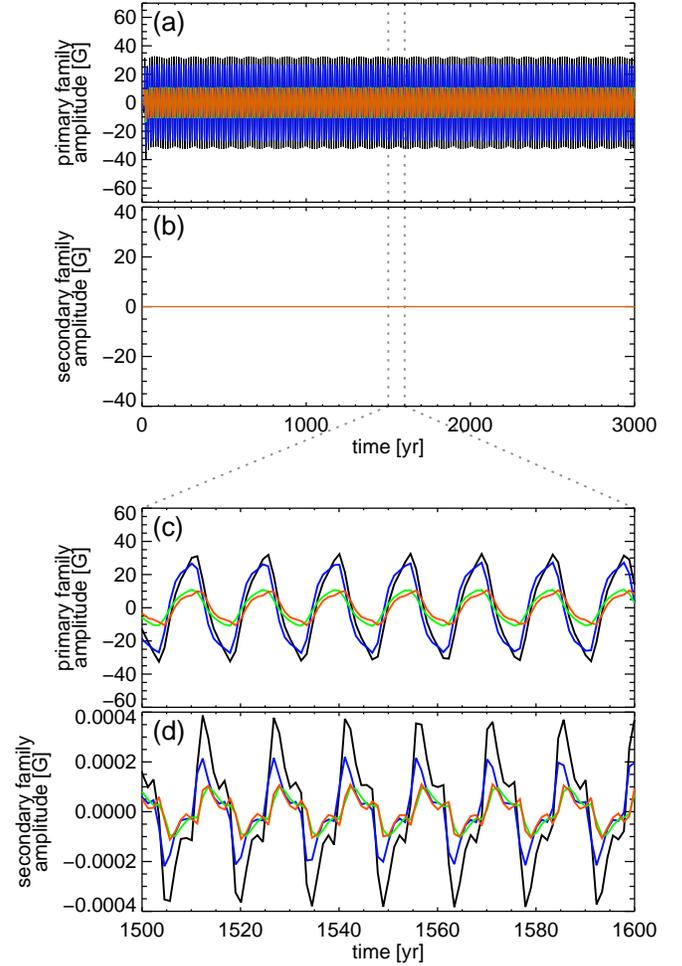}

  \caption{Time evolution of the coefficients of the lowest-order harmonic
    functions of the surface magnetic field (as grouped by primary and
    secondary families) from the same BL dynamo model as shown in
    Fig.~\ref{fig:bl-dipole-butterfly}.  In panel (a) are shown the evolution
    of the first several primary-family coefficients $B_\ell^0$ with $\ell=$1,
    3, 5, and~7 in black, blue, green, and red, respectively.  In panel (b)
    are shown the evolution of the secondary-family coefficients $B_\ell^0$
    with $\ell=$2, 4, 6, and~8, respectively in black, blue, green, and red.
    Panels~(c) and ~(d) show zoomed-in sections of panels~(a) and~(b).}

  \label{fig:bl-dipole}

\end{figure}

To demonstrate these characteristics, we now consider a typical solution of
the standard BL mean-field dynamo as calculated by STELEM.
Figure~\ref{fig:bl-dipole-butterfly} presents the time evolution of the
resulting magnetic field patterns, and is thus analogous to the standard solar
butterfly diagram.  Performing a Legendre transform on the magnetic field
reveals the degrees $\ell$ of the dominant axisymmetric modes.
Figure~\ref{fig:bl-dipole} illustrates that the odd $\ell$ modes from the
primary family dominate over the even ones by about five orders of magnitude
in this model.  This differs significantly from what is observed on the Sun,
where the amplitude of the quadrupole is measured to be about 25\% of the
dipole amplitude for most of the time, becoming dominant only during reversals
[cf., Fig.~\ref{fig:di-vs-quad}(c)].  The behavior of the standard BL model of
Figure~\ref{fig:bl-dipole} arises because of the symmetry characteristics of
the BL dynamo equations.  Because the model was initialized with a dipolar
field, no modes from the secondary family are excited in the standard BL model
shown in Figure~\ref{fig:bl-dipole} because no coupling exists between the
primary and secondary families.

\begin{figure}
  \epsscale{1.0}
  \plotone{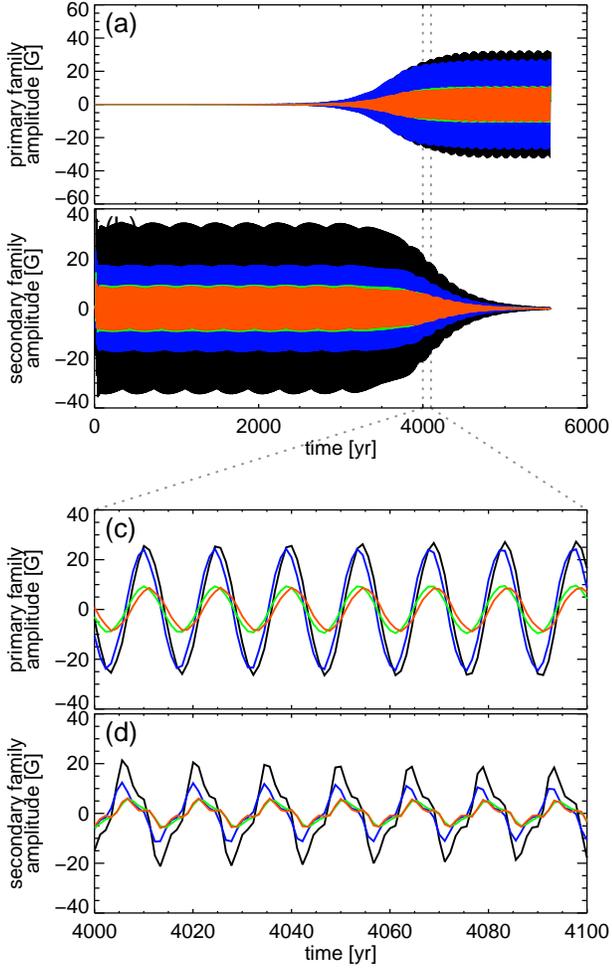}

  \caption{Time evolution of the same BL model as shown
    Fig.~\ref{fig:bl-dipole} (and using the same color scheme), but
    initialized with a quadrupolar magnetic field.}

  \label{fig:bl-quadrupole}

\end{figure}

If instead the calculation were initialized with a quadrupolar field
(belonging to the secondary family), we find that the system eventually
transitions to a state in which the primary-family modes predominate, as shown
in Figure~\ref{fig:bl-quadrupole}.  The growth of the primary-family modes is
due to the presence of a very weak dipole (likely of numerical origin) at the
onset of the simulations.  In these models, the BL source term of
Equation~(\ref{eq:BLsource}) quenches the growth of the magnetic field once a
certain threshold is passed, and as a result the maximum total amplitude of
the magnetic field is capped.  The reason why the primary-family modes are
preferred stems from the fact that the thresholds for dynamo action (based on
the parameter $C_s$ in Equation~[\ref{eqA2}]) are found empirically to be
lower for the dipole than for the quadrupole ($C_s \sim 6.12$ vs.\ $C_s \sim
6.25$), meaning the dipole-like modes have a higher growth rate than the
quadrupole-like modes.  In this model, only briefly during the transition
phase does the model possess a quadrupole of order 25\% of the dipole, as in
the Sun.

Observations of solar photospheric fields, however, indicate that the Sun
excites both families and does not strongly favor members of one family over
the other, a situation that has apparently existed over many centuries.  Even
during the Maunder minimum, evidence suggests that this interval may have been
dominated by a hemispherical dynamo with magnetic activity located primarily
in the southern hemisphere \citep{rib1993}, which can only be formed by a
state in which primary- and secondary-family modes possess nearly equal
amplitudes \citep{tob1997,gal2009}.  Consequently, the solutions presented in
Figures~\ref{fig:bl-dipole} and~\ref{fig:bl-quadrupole}, in which modes from
only one family are preferred, are thus not a satisfactory model of the Sun.

As advocated by \citet{rob1972} and \citet{gub1993} following their
symmetry-based study of the solar dynamo and the induction equation, and more
recently by \citet{nis2008} in their geodynamo simulations, a north-south
asymmetry of the flow field specified in the BL dynamo, or alternatively an
asymmetric poloidal-field source term, may allow the co-existence of both the
primary and secondary families.  To investigate this effect we have performed
two additional BL dynamo calculations, one with a BL source term and one with
a meridional circulation that each generate both symmetric and antisymmetric
fields (introduced via the parameter $\epsilon$ of
Equations~(\ref{eq:asymBLsource}) and~(\ref{eq:asymMC}) of
Appendix~\ref{sec:bleqns}), the combination of which yields asymmetric
magnetic field patterns.

\begin{figure}
  \epsscale{1.0}
  \plotone{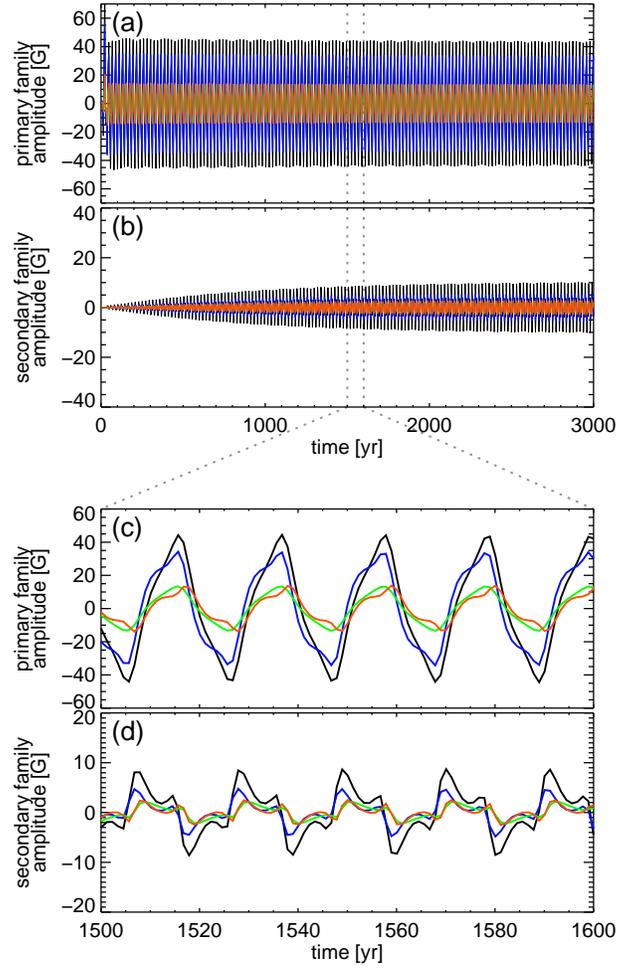}

  \caption{Time evolution of the same BL model as shown in
    Fig.~\ref{fig:bl-dipole} (and using the same color scheme), but with an
    asymmetric BL source term as implemented in Eq.~(\ref{eq:asymBLsource}) by
    setting $\epsilon=10^{-3}$.}

  \label{fig:bl-asymmetric-blterm}

\end{figure}

\begin{figure}
  \epsscale{1.0}
  \plotone{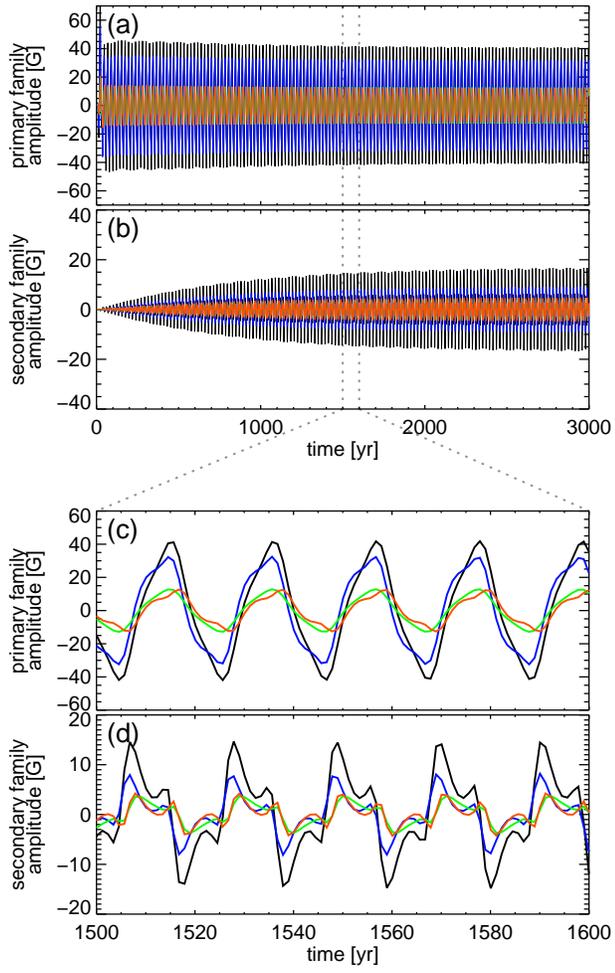}

  \caption{Time evolution of the same BL model as shown in
    Fig.~\ref{fig:bl-dipole} (and using the same color scheme), but with an
    asymmetric meridional flow profile as implemented in Eq.~(\ref{eq:asymMC})
    by setting $\epsilon=10^{-3}$.}

  \label{fig:bl-asymmetric-meridc}

\end{figure}

We have run several dynamo cases, with the antisymmetry arising either in the
BL source term or in the meridional flow profile, and with a range of
amplitudes for the $\epsilon$ parameter from $10^{-4}$ to $10^{-1}$.  All
cases were initialized with a dipolar field.  We find that when $\epsilon$ is
about $10^{-3}$, the modes in secondary family grow until they reach about
35\% of the dominant dipolar mode, as illustrated in
Figures~\ref{fig:bl-asymmetric-blterm} and
\ref{fig:bl-asymmetric-meridc}. This result holds true regardless of whether
the antisymmetry is introduced in the BL source term or in the meridional
circulation profile, with very little difference in the resulting mode
amplitudes.  As expected, using a smaller $\epsilon$ results in solutions
where the primary-family modes dominate, while using a larger $\epsilon$
yields a state where the secondary-family modes are comparable to the
primary-family modes.  Such results may indicate that Sun need only possess a
weak degree of north-south asymmetry in order to behave as it does.

\section{Conclusions} \label{sec:conclusions}

Cycles of magnetic activity in many astrophysical bodies, including the Sun,
Earth, and other stars, are thought to be excited by nonlinear interactions
occurring in their interiors.  Yet in some cases, such as the Sun, the cycles
have approximately regular periods and in others, such as the Earth, there is
no apparent periodicity.  Dynamo theory indicates such a range of behaviors is
expected, and whether the cycles are regular depends on magnetohydrodynamic
parameters that characterize the system, including fluid and magnetic Reynolds
and Rayleigh numbers.  As a consequence, the large-scale appearance of the
magnetic field may provide clues toward the type of dynamo that may be
operating.

In this article, long-term measurements of the solar photospheric magnetic
field are utilized to characterize the waves of dynamo activity that exist
within the interior of the Sun.  Synoptic maps from WSO (dating back to
1976) and MDI (spanning 1995--2010) are used to determine the spherical
harmonic coefficients of the surface magnetic field for the past three sunspot
cycles.  We focus on the apparent interactions between various low-order modes
throughout the past three sunspot cycles, and interpret these trends in the
context of dynamo theory.

The multipolar expansions of the solar field as deduced from WSO and MDI data
indicate that the axial and equatorial dipole modes are out of phase.  During
activity minima, the dipole component of the solar field is generally aligned
with the axis of solar rotation, while the quadrupole component is much
weaker.  During activity maxima, the dipole reverses its polarity with respect
to the rotation axis, and throughout the reversal process there is more energy
in quadrupolar modes than in dipole modes.  During the past three cycles,
these reversals have taken place over a time interval of about 2~yr to 3~yr on
average.  More indirect measures of solar activity, such as the sunspot number
and proxies of the heliospheric field, seem to indicate that such regular
activity cycles have persisted for at least hundreds of years with a period of
approximately 11~yr.  The most recently completed solar cycle (Cycle~23)
lasted for about 13~yr and while unusual, is not unprecedented.  We note in
passing that such modulations of the solar dynamo may be interpreted as a type
of nonlinear interaction between the turbulent alpha effect and the field
and/or flows \citep{tob2002}.

The harmonic modes can also be grouped into primary and secondary families, a
distinction that depends on the north-south symmetry of the various modes.
For example, the axial dipole harmonic is antisymmetric and is a member of the
primary family.  Alternatively, the equatorial dipole and axial quadrupole
modes are both symmetric with respect to the equator and thus are grouped
together in the secondary family.  When the evolution of the mode coefficients
are analyzed in this way, we find that there is a trend for members of the
same family to possess the same phasing, suggesting that modes in the same
family of modes are either excited together and/or are more coupled when
compared with modes of different families.  This coupling is noticeable during
reversals of the solar dipole, as less energy is present in primary-family
modes than in secondary-family modes during these intervals.

The historical record indicates that the geodynamo also undergoes reversals of
its dipole axis (with respect to the rotation axis), but these reversals occur
much more irregularly than in the solar case.  Additionally, the dipole axis
of the terrestrial magnetic field occasionally makes excursions away from the
axis of rotation of the Earth, only to later return without actually
reversing.  An examination of the large-scale harmonic modes of the
geomagnetic field during these intervals indicates that the energy contained
in secondary-family modes was significantly smaller during excursions than
during reversals.  A strong quadrupole during geodynamo reversals is in line
with the solar behavior; there is no parallel with excursions as excursions in
the solar case have not been observed.  Analogous behavior is observed to
occur in the VKS laboratory dynamo with respect to the relative strengths of
the primary and secondary families.

We also examined the coupling of the mode families using a BL mean-field
dynamo model computed using the STELEM code.  Because of the symmetries in the
magnetic induction and in the assumed profiles of the large-scale flow fields
and BL source term, we find that the standard mean-field solar dynamo model
results in a state containing largely members of the primary family.  This is
a result of the dipole (a primary-family mode) being more unstable to dynamo
action than the quadrupole.  With a modest amount of asymmetry, implemented
here either in the meridional flow profile or in the BL source term, we find
from the models that both the primary and secondary families can coexist in
the same model and in the same proportions as in the solar dynamo.  This can
lead to a small lag between the northern and southern hemispheres as is
actually observed on the Sun \citep{dik2007}.

\acknowledgements

M.L.D.\ acknowledges support by the Lockheed Martin SDO/HMI sub-contract
25284100-26967 from Stanford University (through Stanford University prime
contract NAS5-02139).  A.S.B.\ acknowledges financial support by the European
Research Council through grant 207430 STARS2, and by CNRS/INSU via Programme
National Soleil-Terre. A.S.B. is grateful to Alan Title and LMSAL for their
hospitality.  Collection and analysis of WSO and MDI data were supported by
NASA under contracts NNX08AG47G and NNX09AI90G.  The authors thank J.~Aubert,
S.~Fauve, A.~Fournier, M.~Miesch, F.~P{\'e}tr{\'e}lis, E.~Spiegel,
A.~Strugarek, S.~Tobias, J.~Toomre and J.-P.~Zahn for useful discussions.

\facility{{\it Facilities}: WSO, SOHO/MDI}

\appendix

\section{Mean-Field Dynamo Formalism}

In this Appendix, we provide details regarding the Babcock-Leighton mean-field
dynamo models discussed in \S\ref{sec:bl-axisymmetric}.  The pertinent
equations are listed in \S\ref{sec:bleqns} and their symmetry properties are
discussed in \S\ref{sec:symmetry}.

\subsection{Mean-Field Equations}    \label{sec:bleqns}

Here, we briefly list the equations governing the axisymmetric mean-field
dynamo models calculated in \S\ref{sec:bl-axisymmetric}.  A more detailed
explanation can be found in, e.g., \citet{jou2008}.  Following
\citep{mof1978}, the mean-field induction equation is
\begin{equation}
\frac{\partial{\left<\bs{B}\right>}}{\partial t} = \delcr\left(\left<
\bs{V}\right>\bcr\left<\bs{B}\right>\right) + \delcr\left<\bs{v'}\bcr
\bs{b'}\right> - \delcr\left(\eta\delcr\left<
\bs{B}\right>\right), \label{eq:meaninduction}
\end{equation}
where the variables $\left<\bs{B}\right>$ and $\left<\bs{V}\right>$ refer to
the mean parts of the magnetic and velocity fields, and $\bs{v'}$ and
$\bs{b'}$ to their respective fluctuating components.  The function $\eta$ is
the magnetic diffusivity and is not necessarily a constant.  The terms
``mean'' and ``fluctuating'' refer to the fact that a separation of scales has
been performed, such that the mean quantities are computed by averaging over
some appropriate intermediate size scale and the fluctuating quantities are
the residuals.

Working in spherical coordinates $(r,\theta,\phi)$ and under the assumption of
axisymmetry, we perform a poloidal-toroidal decomposition and write the mean
magnetic field $\bs{B}$ and mean velocity field $\bs{V}$ (for clarity the
angle brackets $\langle$ and $\rangle$ are omitted going forward) as 
\begin{align}
\bs{B}(r,\theta,t)&=\delcr\left(A_{\phi}\bs{\hat{e}}_{\phi}\right)+B_{\phi}\bs{\hat{e}}_{\phi}
\label{eq:ptdB} \\ \bs{V}(r,\theta)&=\bs{v_{p}} + \Omega\, r\sin\theta \,
\bs{\hat{e}}_{\phi}, \label{eq:ptdV}
\end{align}
where the poloidal streamfunction $A_\phi(r,\theta,t)$ and toroidal field
$B_\phi(r,\theta,t)$ are used to generate $\bs{B}$.  The velocity field is
time-independent, and is prescribed by profiles for the meridional circulation
$\bs{v_{p}}(r,\theta)$ and differential rotation $\Omega(r,\theta)$.

Rewriting the mean induction equation (\ref{eq:meaninduction}) in terms of
$A_\phi$ and $B_\phi$, we arrive at two coupled partial differential equations
for $A_{\phi}$ and $B_{\phi}$,
\begin{align}
\frac{\partial{A_{\phi}}}{\partial t} &=\frac{\eta}{\eta_{t}}
\left(\nabla^{2}-\frac{1}{\varpi^{2}}\right)A_{\phi}-
R_{e}\frac{\bs{v_p}}{\varpi}\bdot\bdel\left(\varpi
A_{\phi}\right)+C_{s}S \label{eqA2} \\ \frac{\partial {B_{\phi}}}{\partial
  t}&=\frac{\eta}{\eta_{t}}
\left(\nabla^{2}-\frac{1}{\varpi^{2}}\right)B_{\phi}
+\frac{1}{\varpi}\frac{\partial(\varpi B_{\phi})}{\partial r}\frac{\partial
  (\eta/\eta_{t})}{\partial r}
-R_{e}\varpi\bs{v_p}\bdot\bdel\left(\frac{B_{\phi}}{\varpi}\right)-R_{e}B_{\phi}\deldot\bs{v_p}
+C_{\Omega}\varpi\left[\delcr(A_{\phi}\bs{\hat{e}}_{\phi})\right]\bdot\bdel\Omega, \label{eqB2}
\end{align}
where $\varpi=r\sin\theta$.  The contribution to the transport term in the
mean induction equation (\ref{eq:meaninduction}) that arises from the
fluctuating fields, namely the $\delcr\left<\bs{v'}\times\bs{b'}\right>$ term,
is present in the $A_\phi$ equation above and in general is assumed to take a
specific form in terms of the mean magnetic field (cf.,
\citealt{bab1961,lei1969,wany1991a,dik1999,jou2007}).  Here, we use a surface
Babcock-Leighton (BL) term $S(r,\theta,B_{\phi})$ for this purpose which
serves to generate new poloidal field.

Additionally, equations (\ref{eqA2}) and (\ref{eqB2}) have been
nondimensionalized by using $R_\odot$ as the characteristic length scale and
$R_{\odot}^2/\eta_{t}$ as the characteristic time scale, where
$\eta_{t}=10^{11}$ cm$^2$ s$^{-1}$ is representative of the turbulent magnetic
diffusivity in the convective zone.  This rescaling leads to the appearance of
three dimensionless control parameters
$C_{\Omega}=\Omega_{0}R_{\odot}^2/\eta_{t}$, $C_{s}=s_{0}R_{\odot}/\eta_{t}$,
and the Reynolds number $R_{e}=v_{0}R_{\odot}/\eta_{t}$, where $\Omega_{0}$,
$s_{0}$, and $v_{0}$ are respectively the rotation rate and the typical
amplitude of the surface source term and of the meridional flow.

Equations (\ref{eqA2}) and (\ref{eqB2}) are solved with the Stellar Elements
(STELEM; T.~Emonet \& P.~Charbonneau, 1998, priv.\ comm.) code (see Appendix~A
of \citealt{jou2007} for more details) in an annular meridional plane with the
colatitude $\theta$ $\in [0,\pi]$ and the dimensionless radius $r \in
[0.6,1]$, i.e., from slightly below the tachocline ($r\approx 0.7$) up to the
solar surface $R_\odot$.  The STELEM code has been thoroughly tested and
validated via an international mean field dynamo benchmarking process
involving 8 different codes \citep{jou2008}.  At the latitudinal boundaries at
$\theta=0$ and $\theta=\pi$, and at the lower radial boundary at $r=0.6$, both
$A_{\phi}$ and $B_{\phi}$ vanish.  At the upper radial boundary at $r=1$, the
solution is matched to an external potential field.  Usual initial conditions
involve setting a confined dipolar field configuration, i.e. $A_\phi$ is set
to $(\sin\theta) / r^{2}$ in the convective zone and to $0$ below the
tachocline.  To create the simulation shown in Figure~\ref{fig:bl-quadrupole},
the simulation was initialized using a quadrupolar configuration with an
$A_\phi$ of $(3\cos\theta\,\sin\theta) / (2 r^{3})$ in the convection zone.
In both cases, the toroidal field is initialized to 0 everywhere.

The rotation profile used in the series of models discussed in this work
captures many aspects of the true solar angular velocity profile, such as
deduced from helioseismic inversions \citep{tho2003}.  We thus assume
solid-body rotation below $r=0.66$ and a differential rotation above this
tachocline interface as given by the following rotation profile,
\begin{equation}
\Omega(r,\theta)=\Omega_c+\frac{1}{2}\left[1+\text{erf}\left(\frac{2(r-r_c)}{d_1}\right)\right](\Omega_\text{eq}+a_2\cos^2\theta+a_4\cos^4\theta-\Omega_c).
\end{equation}
The parameters $\Omega_\text{eq}=1$, $\Omega_c= 0.93944$, $r_c=0.7$,
$d_1=0.05$, $a_2=-0.136076$ and $a_4=-0.145713$.  With this profile for
$\Omega$, the radial shear is maximal at the tachocline.

We assume that the diffusivity in the envelope $\eta$ is dominated by its
turbulent contribution, whereas in the stable interior $\eta_c \ll \eta_t$. We
smoothly match the two different constant values with an error function which
enables us to continuously transition from $\eta_c$ to $\eta_t$,
\begin{equation}
\eta(r)=\eta_c+\frac{\left(\eta_t-\eta_c\right)}{2}\left[1+\text{erf}\left(\frac{r-r_c}{d}\right)\right],
\label{eqeta}
\end{equation}
with ${\eta_{\rm c}}=10^9 \,\rm cm^2\rm s^{-1}$ and $d=0.03$.

In BL flux-transport dynamo models, the poloidal field owes its origin to the
tilt of magnetic loops emerging at the solar surface. Thus, the source has to
be confined to a thin layer just below the surface and since the process is
fundamentally non-local, the source term depends on the variation of
$B_{\phi}$ at the base of the convection zone. We use the following expression
(which is a slightly modified version of that used in \citealt{jou2007}) in
order to better confine the activity belt to low latitudes:
\begin{equation}
S(r,\theta,B_{\phi})=\frac{1}{2}\left[1+\text{erf}\left(\frac{r-r_2}{d_2}\right)\right]\left[1-\text{erf}\left(\frac{r-1}{d_2}\right)\right]\left[1+\left({\frac{B_\phi(r_c,\theta,t)}{B_0}}\right)^2\right]^{-1}\cos\theta\,\sin^3\theta\,B_\phi(r_c,\theta,t), \label{eq:BLsource}
\end{equation}
where $r_2=0.95$, $d_2=0.01$, $B_0=10^5$G.  In the particular case of an
imposed asymmetry between the north and southern hemisphere
(Fig.~\ref{fig:bl-asymmetric-blterm}) we introduce a modified source term,
modulated by the parameter $\epsilon$, as follows,
\begin{equation}
S_\text{asym}(r,\theta,B_\phi)=\frac{1}{2}\left[1+\text{erf}\left(\frac{r-r_2}{d_2}\right)\right]\left[1-\text{erf}\left(\frac{r-1}{d_2}\right)\right]\left[1+\left({\frac{B_\phi(r_c,\theta,t)}{B_0}}\right)^2\right]^{-1}(\cos\theta
+ \epsilon \sin\theta) \sin^3\theta\,
B_\phi(r_c,\theta,t). \label{eq:asymBLsource}
\end{equation}

In BL flux-transport dynamo models, meridional circulation is used to link the
two sources of the magnetic field, namely the base of the convection zone
(where toroidal field is created via the latitudinal shear) and the solar
surface (where poloidal field is introduced via the BL source term).  In the
series of models discussed in this paper, the meridional circulation is
equatorially symmetric, having one large single cell per hemisphere.  Flows
are directed poleward at the surface and equatorward at depth (as in the Sun),
vanishing at the bottom boundary at $r=0.6$.  The equatorward branch
penetrates slightly beneath the tachocline.  To model the single cell
meridional circulation we consider a stream function with the following
expression \citep{jou2008},
\begin{equation}
\psi(r,\theta)=-\frac{2(r-r_b)^2}{\pi(1-r_b)}\sin\left(\frac{\pi(r-r_b)}{1-r_b}\right)\cos\theta\sin\theta, \label{eq:BLvelstream}
\end{equation}
which gives, through the relation $\bs{v_p}=\delcr(\psi\bs{\hat{e}_\phi})$, the following components of the meridional flow,
\begin{align}
v_{r}&=-\frac{2(1-r_b)}{\pi
  r}\frac{(r-r_b)^2}{(1-r_b)^2}\sin\left(\frac{\pi(r-r_b)}{1-r_b}\right)(3\cos^2\theta-1)
\\ v_{\theta}&=\left[\frac{3r-r_b}{1-r_b}
  \sin\left(\frac{\pi(r-r_b)}{1-r_b}\right)+\frac{r\pi}{1-r_b}\frac{(r-r_b)}{(1-r_b)}
  \cos\left(\frac{\pi(r-r_b)}{1-r_b}\right)\right]\frac{2(1-r_b)}{\pi
  r}\frac{(r-r_b)}{(1-r_b)}\cos\theta\sin\theta,
\end{align}
with $r_b=0.6$.

To introduce asymmetry into the model, an alternative to using the asymmetric
source term of equation~(\ref{eq:asymBLsource}) is to introduce an asymmetry
into the meridional flow profile.  Such an asymmetric meridional flow profile
can be constructed using the following stream function,
\begin{equation}
\psi_\text{asym}(r,\theta)=-\frac{2(r-r_b)^2}{\pi(1-r_b)}\sin\left(\frac{\pi(r-r_b)}{1-r_b}\right)(\cos\theta
+ \epsilon \sin\theta)\sin\theta, \label{eq:asymMC}
\end{equation}
which leads to the following components of the meridional flow,
\begin{align}
v_{r,\text{asym}}&=-\frac{2(1-r_b)}{\pi
  r}\frac{(r-r_b)^2}{(1-r_b)^2}\sin\left(\frac{\pi(r-r_b)}{1-r_b}\right)(3\epsilon\sin\theta\cos\theta+3\cos^2\theta-1) \label{eq:asymVR}
\\
v_{\theta,\text{asym}}&=\left[\frac{3r-r_b}{1-r_b}
  \sin\left(\frac{\pi(r-r_b)}{1-r_b}\right)+\frac{r\pi}{1-r_b}\frac{(r-r_b)}{(1-r_b)}
  \cos\left(\frac{\pi(r-r_b)}{1-r_b}\right)\right]\frac{2(1-r_b)}{\pi
  r}\frac{(r-r_b)}{(1-r_b)}(\cos\theta+ \epsilon \sin\theta)\sin\theta, \label{eq:asymVTH}
\end{align}
again with $r_b=0.6$.

\subsection{Symmetry Considerations}  \label{sec:symmetry}

Following \citet{gub1993} it is straightforward to assess symmetry properties
of various mathematical operators and equations.  We adopt the superscripts
$^A$ and $^S$ to indicate whether scalars or vectors are antisymmetric or
symmetric across the equator, respectively.  For example, products between a
scalar and a vector of the form $a \bs{F}=\bs{G}$, where $a$ and $\bs{F}$ are
of like symmetry, yield a symmetric result ($a^S\bs{F}^S\rightarrow\bs{G}^S$
and $a^A\bs{F}^A\rightarrow\bs{G}^S$), whereas products between quantities of
differing symmetries are antisymmetric ($a^A\bs{F}^S\rightarrow\bs{G}^A$ and
$a^S\bs{F}^A\rightarrow\bs{G}^A$).  For the vector cross product
$\bs{F}\bcr\bs{G}=\bs{H}$, when the two vectors $\bs{F}$ and $\bs{G}$ have the
same symmetry properties the result will be antisymmetric
($\bs{F}^S\bcr\bs{G}^S\rightarrow\bs{H}^A$ and
$\bs{F}^A\bcr\bs{G}^A\rightarrow\bs{H}^A$), while the cross product between
two vectors having opposing symmetries will yield a symmetric result
($\bs{F}^A\bcr\bs{G}^S\rightarrow\bs{H}^S$).  Additionally, the curl operator
reverses symmetry ($\delcr\bs{G}^A\rightarrow\bs{H}^S$ and
$\delcr\bs{G}^S\rightarrow\bs{H}^A$), while the Laplacian operator preserves
symmetry ($\nabla^2\bs{F}^S\rightarrow\bs{H}^S$ and
$\nabla^2\bs{F}^A\rightarrow\bs{H}^A$).

With these properties established, the analysis of the symmetry properties of
the magnetic induction equation,
\begin{equation}
\frac{\partial\bs{B}}{\partial t}=\delcr(\bs{V}\bcr\bs{B}) +
\eta\nabla^2\bs{B}, \label{eq:fullinduct}
\end{equation}
follows in a straightforward manner.  For cases possessing a symmetric
velocity field $\bs{V}^S$ with respect to the equator, both terms on the
right-hand side of equation~(\ref{eq:fullinduct}) preserve the symmetry of
$\bs{B}$.  Thus, a dynamo having only a symmetric field $\bs{B}^S$ will remain
symmetric over time, since both the transport term and the diffusion term of
equation~(\ref{eq:fullinduct}) generate symmetric field only.  Likewise, a
dynamo possessing an antisymmetric field $\bs{B}^A$ will preserve its
antisymmetry over time.  Because equation~(\ref{eq:fullinduct}) is linear in
$\bs{B}$, it follows that a magnetic field possessing mixed symmetry in the
midst of a symmetric velocity field can be considered to be operating two
independent, noninteracting dynamos: one that is symmetric and one that is
antisymmetric.  

However, in cases with an antisymmetric velocity field $\bs{V}^A$, the
transport term on the right-hand side of equation~(\ref{eq:fullinduct})
provides a mechanism by which the symmetric and antisymmetric modes of
$\bs{B}$ can couple.  This coupling arises because an initially symmetric
field $\bs{B}^S$ will first generate an antisymmetric field that in turn leads
to the presence of both antisymmetric and symmetric fields, according to the
right-hand side of equation~(\ref{eq:fullinduct}).  Analogously, initializing
with a purely antisymmetric field $\bs{B}^A$ will generate fields of mixed
symmetry over time.

This analysis procedure can further be applied to the mean-field induction
equation.  For example an analysis of equation~(\ref{eq:meaninduction}) with
an $\alpha$-$\omega$ dynamo (i.e., where $\left<\bs{v}'\bcr\bs{b}'\right>$ is
set to $\alpha\left<\bs{B}\right>$),
\begin{equation}
\frac{\partial{\left<\bs{B}\right>}}{\partial t} = \delcr\left(
\left<\bs{V}\right> \bcr \left<\bs{B}\right> + \alpha \left<\bs{B}\right>
\right) - \eta \nabla^2 \left(\left<\bs{B}\right>
\right) \label{eq:alphaomega}
\end{equation}
indicates that, for an assumed symmetric mean velocity field
$\left<\bs{V}\right>^S$ and an antisymmetric alpha effect $\alpha^A$ (which is
the natural outcome of helical turbulence in a rotating fluid), such a
mean-field dynamo will preserve the symmetry (or antisymmetry) of the initial
fields.  Hence, as pointed out by \citet{rob1972} and \citet{mcf1991}, a
symmetric mean velocity field and an antisymmetric alpha effect do not couple
magnetic field modes belonging to different families.  Alternatively, if
instead an antisymmetric mean flow $\left<\bs{V}\right>^A$ or a symmetric
alpha effect $\alpha^S$ are considered, this now enables a coupling between
symmetric and antisymmetric mean fields.

In a similar vein, the BL equations~(\ref{eqA2}) and~(\ref{eqB2}), which are
determined by performing the poloidal-toroidal decomposition on
equation~(\ref{eq:alphaomega}), can also be analyzed for symmetry.  It is
important to note that, by equation~(\ref{eq:ptdB}), the poloidal
streamfunction $\left<A_\phi\right>$ has a symmetry opposite to that of the
mean magnetic field $\left<\bs{B}\right>$ it generates (and thus also to the
corresponding toroidal field $\left<B_\phi\right>$).  We established above
that the diffusion term in the mean-field equation~(\ref{eq:alphaomega})
preserves the symmetry of $\left<\bs{B}\right>$, and so it follows that the
corresponding diffusion terms in the BL dynamo equations~(\ref{eqA2})
and~(\ref{eqB2}) will serve to preserve the symmetries $A_\phi$ and $B_\phi$.
Likewise, because the large-scale transport term $\delcr\left(
\left<\bs{V}\right> \bcr \left<\bs{B}\right>\right)$ term in
equation~(\ref{eq:alphaomega}) preserves the symmetry of $\left<\bs{B}\right>$
whenever $\left<\bs{V}\right>$ is symmetric, it follows that the analogous
terms in the equations~(\ref{eqA2}) and~(\ref{eqB2}) also preserve the
symmetry of the system as long as the imposed velocity field is symmetric.
For the BL dynamo considered here, an antisymmetric poloidal velocity
streamfunction $\psi^A$, as in equation~(\ref{eq:BLvelstream}), yields a
symmetric meridional flow profile $\bs{v_p}^S$, since
$\bs{v_p}=\delcr(\psi\bs{\hat{e}_\phi})$, which in turn gives a symmetric mean
velocity $\left<\bs{V}\right>^S$ from equation~(\ref{eq:ptdV}).  Therefore,
the imposed velocity field as defined by equations~(\ref{eq:ptdV})
and~(\ref{eq:BLvelstream}) will preserve the symmetries of $A_\phi$ and
$B_\phi$.  Lastly, the source term $S$ as defined by
equation~(\ref{eq:BLsource}) also preserves the symmetry of $A_\phi$, since it
is comprised of a series of symmetric coefficients multiplied by $\cos\theta\,
B_\phi$.  Therefore, an antisymmetric toroidal field implies a symmetric
source term that in turn serves to preserve the symmetry of $A_\phi$ (and thus
$\left<\bs{B}\right>$), and the same is true when the toroidal field is
symmetric.

For these reasons, the dynamo whose characteristics are illustrated in
Figure~\ref{fig:bl-dipole}, which was initialized with a dipolar field (which
is antisymmetric), preserves its antisymmetry with time since all of the terms
in equations~(\ref{eqA2}) and~(\ref{eqB2}) preserve the initial symmetry.
Indeed, the amplitude of the secondary-family modes (which are symmetric)
remain low in this model, as shown in Figures~\ref{fig:bl-dipole}(b) and~(d).
In the dynamos whose characteristics are displayed in
Figures~\ref{fig:bl-asymmetric-blterm} or~\ref{fig:bl-asymmetric-meridc}, this
effect is responsible for the growth of symmetric mean fields, even though
both models were initialized with the same antisymmetric mean magnetic field.

It is therefore a direct outcome of symmetry considerations that in standard
mean-field dynamo models either one or the other families of magnetic fields
is excited.  In the experiments discussed earlier in
\S\ref{sec:bl-axisymmetric}, we controlled the degree to which the symmetries
were mixed via the parameter $\epsilon$ in equation~(\ref{eq:asymBLsource})
and in equations~(\ref{eq:asymVR}) and~(\ref{eq:asymVTH}), which led to the
dynamos illustrated in Figures~\ref{fig:bl-asymmetric-blterm}
or~\ref{fig:bl-asymmetric-meridc}, respectively.  In both cases, $\epsilon$
was chosen to yield a dynamo where the end state contained secondary-family
amplitudes of about 25\%, as is observed on the Sun; other choices of
$\epsilon$ will lead to end states with different ratios.

\bibliographystyle{apj}


\end{document}